\documentclass[twocolumn,prb,showpacs,preprintnumbers,amsmath,amssymb,superscriptaddress]{revtex4}
\usepackage{graphicx}% Include figure files
\usepackage{dcolumn}% Align table columns on decimal point
\usepackage{color}% To mark revisions.  Delete this command and all \textcolor commands before submission
\usepackage{bm}% bold math
\usepackage{epsfig}
\begin{document}
\draft
\preprint{HEP/123-qed}
\title{Designing Dirac points in two-dimensional lattices}
\author{Kenichi Asano}
\affiliation{Department of Physics, Osaka University, Toyonaka, Osaka 560-0043, Japan}
%\footnote{\vspace*{-10mm} electric address: asano@phys.sci.osaka-u.ac.jp} 
\author{Chisa Hotta}
%\email[Electronic address: ]{chisa@cc.kyoto-su.ac.jp}
\affiliation{Department of Physics, Faculty of Science, Kyoto Sangyo University, Kyoto 603-8555, Japan}
\date{Submitted: October 14, 2010; Accepted: April 2010}
%%{submitted in }
%*%*%*%*%*%*%*%*%*%*%*%*%*%*%*%*%*%*%*%*%*%*%*%*
\begin{abstract}
We present a framework to elucidate the existence of {\it accidental} contacts of energy bands, particularly those called Dirac points which are the point contacts with linear energy dispersions in their vicinity. 
A generalized von-Neumann-Wigner theorem we propose here gives the number of constraints on the lattice necessary to have contacts without fine tuning of lattice parameters. 
By counting this number, one could quest for the candidate of Dirac systems {\it without solving the secular equation}. 
The constraints can be provided by any kinds of symmetry present in the system. 
The theory also enables the analytical determination of $\bm k$-point having {\it accidental} contact by selectively picking up only the degenerate solution of the secular equation. 
By using these frameworks, we demonstrate that the Dirac points are feasible in various two-dimensional lattices, e.g. 
the anisotropic Kagom\'e lattice under inversion symmetry is found to have contacts over the whole lattice parameter space. 
Spin-dependent cases, such as the spin-density-wave state in LaOFeAs with reflection symmetry, are also dealt with in the present scheme. 
\end{abstract}
%*%*%*%*%*%*%*%*%*%*%*%*%*%*%*%*%*%*%*%*%*%*%*%*
\vskip2pc
\maketitle
%*%*%*%*%*%*%*%*%*%*%*%*%*%*%*%*%*%*%*%*%*%*%*%*
\section{introduction}\label{sec:intro}
An issue regarding the contacts of energy bands has long been studied from the early stage of solid state physics\cite{herring37}. 
The one currently attracting wide interest is called Dirac point which is characterized by the linear splitting of energy bands in its vicinity.
This distinguishing band structure leads to the exotic phenomena in the electronic transport, orbital diamagnetism, etc\cite{ando07}.
The Dirac points have been observed and studied in numbers of materials: graphene\cite{ando07,Novoselov04,Novoselov05,Zhang05}, organic solids such as $\alpha$-ET$_2$I$_3$\cite{tajima02,akobayashi07}, LaOFeAs\cite{LaOFeAs,ran09} in its spin-density-wave state, and those with spin-orbit interactions\cite{topo}. 
The ones in graphene are located at K and K'-points at the corners of the hexagonal Brillouin zone\cite{Wallace47,Lomer1955}. 
However, in $\alpha$-ET$_2$I$_3$, they fall on some general $\bm k$-points inside the Brillouin zone \cite{tajima02,akobayashi07,kino07}. 
In two-dimensional(2D) models, their presence are shown in several examples, such as the isotropic\cite{Wallace47} and anisotropic\cite{wunsch08} honeycomb lattices, the anisotropic square lattice under the presence of a $\pi$-flux\cite{affleck88}, the non-Bravais anisotropic square ones\cite{tmori}, and the isotropic Kagom\'e lattice\cite{Guo09}, etc. 
The first-principle\cite{kino07}, semi-empirical calculations\cite{rkondo09}, and perturbative calculations\cite{katayama09} are used to show the existence of Dirac points in $\alpha$-ET$_2$I$_3$. 
However, almost all these previous studies focus only on the specific and detailed models or materials, and lack the general viewpoint of {\it how to design the systems with Dirac points in their energy bands (Dirac systems)}. 
\par
The contact, or equivalently, degeneracy of energy bands at the same $\bm k$-point, is classified into the {\it essential} or {\it accidental} (nonessential) ones, according to whether or not we can specify that $\bm k$-point {\it in advance} without solving secular equation.
By definition, the essential degeneracy {\it is forced to take place at special $\bm k$-points}, while the accidental ones can {\it sometimes occur at general $\bm k$-points}.
Thus, the Dirac points found in $\alpha$-ET$_2$I$_3$ is clearly {\it accidental}.
Although the general theory on the essential degeneracy is well established, that on accidental one has been only poorly explored. 
Albeit, the demand for the latter theory is developed recently, with a need to design Dirac systems.  
\par
In the present paper, we propose a feasibility (generalized von-Neumann) theorem, which provides a general treatment on the accidental degeneracy of energy bands. 
On the basis of this theorem, one could explore the candidates of Dirac systems only by counting the number of constraints on the lattice, {\it without solving the secular equation}. 
Our framework also provides a practical procedure to search for the $\bm k$-points at which the accidental degeneracy takes place (see the introductory part of \S\ref{sec:Dirac}). 
By this procedure, one can {\it selectively} pick up only the degenerate solutions of the secular equation, which is indispensable in the analysis of the accidental degeneracy. 
\par
In \S\ref{sec:general}, we present a feasibility theorem in detail. 
In \S\ref{sec:symmetry}, the relation between the symmetry of the system and the number of constraints is discussed. 
Then, we demonstrate in \S\ref{sec:Dirac} that Dirac systems are easily designed on various 2D lattices. 
In fact, the present study lead to the rediscovery of another Dirac systems in multi-band two-dimensional organic crystal, (DIEDO)$_2X$ ($X$=Cl,Br) \cite{imakubo06}. 
Those who are not interested in the details of the formulation can first go through \S\ref{sec:preliminary} and then directly apply the practical procedure explained in the first part of \S\ref{sec:Dirac}. 
\par
%*%*%*%*%*%*%*%*%*%*%*%*%*%*%*%*%*%*%*%*%*%*%*%*
\section{General Consideration}\label{sec:general}
\subsection{Preliminary definitions}\label{sec:preliminary}
We consider a lattice with $n_{\rm s}$ orbitals per primitive unit cell.
Our primary concern is the non-interacting system, while interaction effects can be treated via the one-body approximation. 
The effect of external magnetic field is also not considered, so that the system is invariant under the time-reversal operation. 
However, our formalism can be easily extended to the systems in a uniform magnetic field, by the introduction of an extended magnetic unit cell with an integer magnetic flux quanta and a corresponding folded magnetic Brillouin zone.
Here, the spin degrees of freedom is also formally neglected, 
whereas we can take it into account by regarding up- and down-spins as different orbitals as mentioned in \S\ref{sec:spin}, which simply doubles the number of the orbitals. 
\par
Following Ref.~\onlinecite{herring37}, we are using the term {\it contact} to denote the degeneracy of energy bands at the same $\bm k$-point.
Let the contact be called $n_{\rm d}$-dimensional, when it takes place on a $n_{\rm d}$-dimensional manifold in the $\bm k$-space.
In particular, the zero- and one-dimensional ones are named  {\it point contact} and  {\it line contact}, respectively.
The point contact at $\bm k=\bm k_0$ is designated as {\it Dirac point}, when two bands split linearly in energy for any small $\delta\bm k=\bm k-\bm k_0$ and form an elliptic Dirac cone.
Besides, we can also consider the situation for $n_{\rm s}>2$, where three or more bands are touched together at the same $\bm k$-point.
It is expressed as {\it $m$-fold contact}, where $m$ is the number of bands touched. 
\par
The lattice system is characterized by a set of parameters, which we call {\it lattice parameters}, $\bm t=(t_1,t_2,\cdots)$.
For example, in the tight binding models, $\bm t$ indicates the transfer integrals, the energy levels of the sites, and the spatial coordinates of sites, and so on. 
In the first principle calculations, $\bm t$ denotes the Fourier components of the self-consistent lattice potential. 
If the contact takes place in finite and connected region in the lattice parameter space, the contacts are called {\it feasible}. 
By contrast, if its presence needs fine tuning of the lattice parameters\cite{vanishing}, or equivalently, if it occurs only at an isolated point in the lattice parameter space, the contact is considered to be {\it unfeasible}. 
The latter is unworthy of attention, because it is too fragile to be observed in the realistic materials. 
\par
As will be eventually shown in \S\ref{sec:GvNW-theorem}, in order to explore the candidates of lattices with {\it feasible} contacts, one has only to count the number of constraints on the lattice parameters, $n_{\rm c}$, which are often provided by symmetries. 
Actually, the dimension, $n_{\rm d}$, of the feasible $m$-fold contact should satisfy the {\it feasibility (or generalized von-Neumann-Wigner) condition},
\begin{equation}
n_{\rm d}=n_{\rm u}-m^2+1+n_{\rm c}\ge 0,
\label{eq:n_d}
\end{equation}
where $n_{\rm u}$ denote the number of unknown variables used in searching for the contact.
The number of constraints and unknowns satisfy $0\le n_{\rm c}\le m^2-1$, and $0\le n_{\rm u}\le d$, respectively, where $d$ denotes the spatial dimension.
When one searches {\it general $\bm k$-points} on the Brillouin zone, the number of unknowns is given as $n_{\rm u}=d$.
Sometimes, the number of constraint, $n_{\rm c}$, increases at {\it special $\bm k$-points}.
In this case, the number of unknowns, $n_{\rm u}$, should be redefined as the dimension of manifold of these special $\bm k$-points. 
As an example, let us consider a two-dimensional lattice $(d=2)$ with a reflection symmetry. 
The $\bm k$-points on the symmetry axis must be distinguished from the general $\bm k$-points, since one finds larger number of constraints. 
The number of unknowns is reduced to $n_{\rm u}=1$ there. 
\par
In our context, the contact is considered to be {\it essential}, when there exist the maximum number ($n_{\rm c}=m^2-1$) of constraints on some special $\bm k$-points. 
Actually, one {\it always } finds $n_{\rm u}$-dimensional feasible contact on these special $\bm k$-points.
Otherwise, the contact is accidental, and thus we need to search over the $\bm k$-space under the properly chosen lattice parameters. 
It is noteworthy that the accidental contact can be feasible on the general $\bm k$-points in contrast to the essential one. 
\par
Among the feasible contacts, the one at $\bm k=\bm k_0$ is considered to be {\it stable}, 
if we can find a new contact in the neighborhood of $\bm k=\bm k_0$ after {\it any} infinitesimally small variation of lattice parameters, $\bm t\rightarrow\bm t+\delta\bm t$ under the above mentioned constraints.
Conversely, contacts become {\it unstable}, just when they are created or destructed by some variation of lattice parameters. 
More intuitively, the stable contacts are realized inside the lattice parameter region showing contacts, while the unstable ones on its edge.
\par
%*%*%*%*%*%*%*%*%*%*%*%*%*%*%*%*%*%*%*%*%*%*%*%*
\subsection{Contacts in two-band systems in 2D}\label{sec:Dirac-2band}
Let us begin with the two-band systems ($n_{\rm s}=2$) in two dimension.
The properties of point contacts in these systems have already been investigated in detail, 
and most of such studies depend on the topological arguments\cite{hatsugai10,ran09}. 
In this section, we present some elementary explanation as an introduction to the multi-band cases.
\par
The Hamiltonian at the Bloch wave vector, $\bm k=(k_1,k_2)$, is expressed by a $2\times 2$ Hermite matrix. 
Following the notations of Ref.~\onlinecite{hatsugai10}, we expand it as 
\begin{equation}
\hat H({\bm k})=E_{\rm 0}(\bm k)\hat I+\bm R(\bm k) \cdot\hat{\bm\sigma}, 
\label{eq:ham2}
\end{equation}
where $E_{\rm 0}(\bm k)$ and $\bm R(\bm k)\!=\!(R_1(\bm k),R_2(\bm k),R_3(\bm k))$ are real functions of $\bm k$, and $\hat{\bm\sigma}=(\hat\sigma_1,\hat\sigma_2,\hat\sigma_3)$ are the Pauli matrices defined by
\begin{equation}
\hat\sigma_1=\left(
\begin{array}{cc}
0&1\\
1&0
\end{array}
\right),\ 
\hat\sigma_2=\left(
\begin{array}{cc}
0&-i\\
i&0
\end{array}
\right),\ 
\hat\sigma_3=\left(
\begin{array}{cc}
1&0\\
0&-1
\end{array}
\right).
\end{equation}
Such an expansion is always possible, because the unit matrix, $\hat I$, and the Pauli matrices form a complete basis for the $2\times 2$ Hermite matrices.
Regarding the second term in Eq.~(\ref{eq:ham2}) as a fictitious Zeeman splitting, we immediately obtain the eigenenergies as
\begin{equation}
\epsilon_{\pm}(\bm k)=E_{\rm 0}(\bm k) \pm |\bm R(\bm k)|, 
\end{equation}
which shows that a contact takes place at $\bm k=\bm k_0$, where 
\begin{equation}
\bm R(\bm k)=\bm 0
\label{eq:cond}
\end{equation}
is satisfied, and its energy is given as $\epsilon_0=E_0(\bm k_0)$.
It is intrinsic that three conditions ($R_1=R_2=R_3=0$) must be simultaneously fulfilled to have a degeneracy, 
which is known as von-Neumann-Wigner theorem\cite{neumannwigner29,Landau}. 
\par
Here, we consider only the point contact located at a general $\bm k$-point.
Then, we have two unknowns, $k_1$ and $k_2$, and thus Eq.~(\ref{eq:cond}) is overdetermined. 
In order to make the point contact at $\bm k=\bm k_0$ feasible, a constraint is required, which reduce the number of conditions by one. 
In the vicinity of $\bm k=\bm k_0$, this constraint should be expressed as
\begin{equation}
\bm s \cdot \bm R(\bm k)=0,
\label{eq:nR}
\end{equation}
within the linear order of $\bm R(\bm k)$, where $\bm s$ is a nonzero three-dimensional (3D) vector.
As will be mentioned in \S \ref{sec:symmetry} (in the context of Eq.(\ref{eq:nR2}) for multi-band systems), this constraint is usually attributable to the symmetries of the system. 
\par
If a point contact arise at $\bm k=\bm k_0$, the energy bands are expanded within the linear order of $\delta\bm k=\bm k-\bm k_0$ as
\begin{equation}
\epsilon_\pm(\bm k_0+\delta\bm k)=\epsilon_0+\bm A\cdot\delta\bm k\pm\left|\bm X\delta k_1+\bm Y\delta k_2\right|, 
\end{equation}
with
\begin{equation}
\bm X=\nabla_{k_1}\bm R|_{\bm k=\bm k_0},\ \bm Y=\nabla_{k_2}\bm R|_{\bm k=\bm k_0},\ \bm A=\nabla_{\bm k} E_0|_{\bm k=\bm k_0}.\label{eq:XYA}
\end{equation}
This contact is identified as a Dirac point, when $\epsilon_\pm$ form an elliptic Dirac cone in the vicinity of $\bm k=\bm k_0$\cite{Landau}, or equivalently, if two vectors, $\bm X$ and $\bm Y$, are linearly independent. 
The Dirac cone tilts, if $\bm A \ne \bm 0$\cite{Goerbig08}. 
\par
A feasible point contact, which is realized at $\bm k=\bm k_0$ for the lattice parameter, $\bm t=\bm t_0$, is classified to {\it stable} and {\it unstable} ones, whether or not we can find a new point contact in the vicinity of the original one after the infinitesimally small change of lattice parameter, $\bm t=\bm t_0+\delta\bm t$, under the constraint of Eq.~(\ref{eq:nR}).
Within the linear order of $\delta\bm k$ and $\delta\bm t$, we obtain
\begin{align}
\bm R(\bm k_0+\delta\bm k,\bm t_0+\delta\bm t)&=\bm R(\bm k_0+\delta\bm k,\bm t_0+\delta\bm t)-\bm R(\bm k_0,\bm t_0)\notag\\
&=\bm X\delta k_1+\bm Y\delta k_2+\delta\bm R_{\delta\bm t},
\end{align}
where two vectors, $\bm X$ and $\bm Y$, defined in Eq.~(\ref{eq:XYA}) are evaluated at $\bm t=\bm t_0$, and another vector is introduced as
\begin{equation}
\delta \bm R_{\delta \bm t}=(\delta\bm t\cdot\nabla_{\bm t})\bm R|_{\bm k=\bm k_0,\bm t=\bm t_0}. 
\end{equation}
Due to the constraint of Eq.~(\ref{eq:nR}), three vectors, $\bm X$, $\bm Y$, and $\bm R_{\delta \bm t}$ are coplanar, satisfying $\bm s\cdot\bm X=\bm s\cdot\bm Y=\bm s\cdot\delta \bm R_{\delta \bm t}=0$.
If two vectors, $\bm X$ and $\bm Y$ are linearly independent, the feasible point contact is stable. 
Actually, the equation, $\bm R(\bm k_0+\delta\bm k,\bm t_0+\delta\bm t)=\bm 0$, is solvable as a linear equation of $\delta\bm k$, and a new point contact is found.
In other words, Dirac points are, by definition, stable.
\par
\begin{figure}[tbp]
\begin{center}
\includegraphics[width=7.5cm]{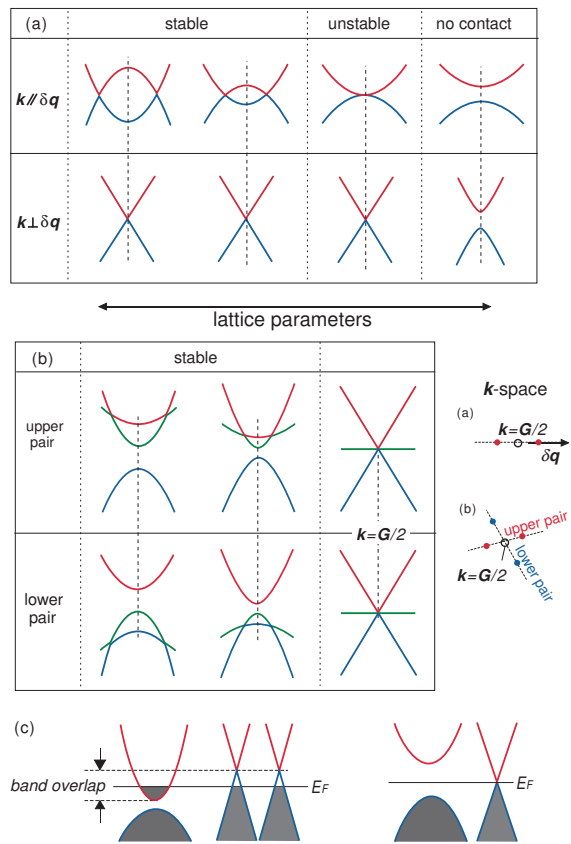}
\end{center}
\caption{(Color online) 
(a) Schematic illustration of the merging of two Dirac points along the direction parallel and perpendicular to $\delta \bm q$ (direction along which the Dirac points merge). 
(b) Merging of four Dirac points consisting of upper- and lower-pairs along the direction parallel to $\delta \bm q$'s of each pair. 
(c) Sketches of energy bands when the band overlap is present (left panel) and absent (right panel). 
}
\label{fig:f1}
\end{figure}

Conversely, whenever the feasible contact is unstable, or equivalently, just created by some lattice parameter change, $\bm X$ and $\bm Y$ are linearly dependent, and thus there exists a vector, $\delta\bm q\ne\bm 0$, satisfying $\bm X\delta q_1+\bm Y\delta q_2=\bm 0$.
Thus, the band splitting $\Delta(\delta\bm k)=\epsilon_+(\bm k_0+\delta\bm k)-\epsilon_-(\bm k_0+\delta\bm k)=2|\bm R(\bm k_0+\delta\bm k)|$ show a quadratic dependence on $\delta\bm k$ when $\delta\bm k$ is parallel to $\delta\bm q$.
In other words, the unstable point contact is given as a doubly degenerate solution of Eq.~(\ref{eq:cond}), i.e., as the consequence of the merging of two Dirac points, as depicted schematically in Fig.~\ref{fig:f1}(a).
The pair creation and destruction of Dirac points implies that there are always even number of Dirac points.
This is nothing but Nielssen-Ninomiya (Fermion doubling) theorem in 2D lattices\cite{ninomiya81}. 
\par
Under the time-reversal symmetry, emergence of even number of Dirac points becomes trivial as discussed in Ref.~\onlinecite{montambaux09}: a Dirac point at $\bm k=\bm k_0$ is always accompanied by its pair at $\bm k=-\bm k_0$.
If there is only a single pair of Dirac points at $\bm k=\pm\bm k_0$, they can be merged only at the special $\bm k$-points invariant under the transformation $\bm k\leftrightarrow -\bm k$, which are given as $\bm k=\bm G/2$ with a reciprocal lattice vector, $\bm G$.
\par
%*%*%*%*%*%*%*%*%*%*%*%*%*%*%*%*%*%*%*%*%*%*%*%*
%*%*%*%*%*%*%*%*%*%*%*%*%*%*%*%*%*%*%*%*%*%*%*%*
%
%*%*%*%*%*%*%*%*%*%*%*%*%*%*%*%*%*%*%*%*%*%*%*%*
\subsection{Generalized von-Neumann-Wigner theorem}\label{sec:GvNW-theorem}
Most of recent studies on the point contacts mainly focus on ones discussed in the previous section, i.e., the ordinary (two-fold) point contact in 2D lattices with two bands, which is located at a general $\bm k$-point. 
In this section, we develop a formalism for the general cases; the $m$-fold contacts in $d$-dimensional systems with $n_{\rm s}$ bands ($m\le n_{\rm s}$), which are located at general or special $\bm k$-points. 
We call it {\it generalized von-Neumann-Wigner theorem}. 
\par
Let us begin with the case of $m=n_{\rm s}$.
The Hamiltonian at the Bloch wave vector, $\bm k=(k_1,k_2,\cdots,k_d)$, can be represented as a $m\times m$ matrix. 
We expand it in the same form as Eq.~(\ref{eq:ham2}):
\begin{equation}
\hat H(\bm k)=E_0(\bm k)\hat I+\bm R(\bm k)\cdot\hat{\bm\sigma}^{(m)},
\label{eq:ham3}
\end{equation}
where $E_0(\bm k)$ and the $(m^2-1)$-dimensional vector $\bm R(\bm k)=(R_1(\bm k),R_2(\bm k),\cdots,R_{m^2-1}(\bm k))$ are real, and the traceless Hermite matrices, $\bm{\hat\sigma}^{(m)}=(\hat\sigma_1,\hat\sigma_2,\cdots,\hat\sigma_{m^2-1})$, denote the generalized Pauli matrices, which span the whole linear space of the $m\times m$ Hermite matrices.
For example, in case of $m=3$, they can be chosen as the Gell-Mann matrices:
\begin{align}
\hat\sigma_1&\!\!=\!\!\left(
\begin{array}{ccc}
0&1&0\\1&0&0\\0&0&0
\end{array}\right),\;
\hat\sigma_2\!\!=\!\!\left(
\begin{array}{ccc}
0&-i&0\\i&0&0\\0&0&0
\end{array}\right),\;
\hat\sigma_3\!\!=\!\!\left(
\begin{array}{ccc}
1&0&0\\0&-1&0\\0&0&0
\end{array}\right),\notag\\
\hat\sigma_4&\!\!=\!\!\left(
\begin{array}{ccc}
0&0&1\\0&0&0\\1&0&0
\end{array}\right),\;
\hat\sigma_5\!\!=\!\!\left(
\begin{array}{ccc}
0&0&-i\\0&0&0\\i&0&0
\end{array}\right),\;
\hat\sigma_6\!\!=\!\!\left(
\begin{array}{ccc}
0&0&0\\0&0&1\\0&1&0
\end{array}\right),\notag\\
\hat\sigma_7&\!\!=\!\!\left(
\begin{array}{ccc}
0&0&0\\0&0&-i\\0&i&0
\end{array}\right),\;
\hat\sigma_8\!\!=\!\!\frac 1{\sqrt 3}\left(
\begin{array}{ccc}
1&0&0\\0&1&0\\0&0&-2
\end{array}\right). 
\label{eq:gellmann}
\end{align}
If we find an $m$-fold contact of energy $\epsilon_0$ at $\bm k=\bm k_0$, there exists a unitary matrix $\hat U$, which satisfies $\hat U\hat H(\bm k_0)\hat U^{-1}=\epsilon_0\hat I$.
This implies $\hat H(\bm k_0)=\epsilon_0\hat I$: $(m^2-1)$ conditions, $\bm R(\bm k)=\bm 0$, should be satisfied at $\bm k=\bm k_0$, and the energy of the contact is given as $\epsilon_0=E_0(\bm k_0)$.
\par
Now, let us proceed to the case of $n_{\rm s}> m$. 
Instead of direct diagonalization of $n_{\rm s}\times n_{\rm s}$ Hamiltonian, $\hat H(\bm k)$, one could renormalize it into smaller dimension following the formalism of Brillouin and Wigner\cite{br-wigner}.
Let us divide the $n_{\rm s}$-dimensional Hilbert space $S$ into $m$-dimensional subspace $S_{\rm A}$ and $(n_{\rm s}-m)$-dimensional subspace $S_{\rm B}$.
Then, the Hamiltonian matrix and its resolvent read
\begin{align}
\hat H({\bm k})&=
\left(
\begin{array}{cc}
\hat H_{\rm AA}(\bm k)   &  \hat H_{\rm AB}(\bm k) \\
\hat H_{\rm BA}(\bm k)   &  \hat H_{\rm BB}(\bm k) 
\end{array}
\right)\\
\hat G({\bm k},\epsilon)&=\left(\epsilon-\hat H(\bm k)\right)^{-1}\!\!\!=
\left(
\begin{array}{cc}
\hat G_{\rm AA}(\bm k)   &  \hat G_{\rm AB}(\bm k) \\
\hat G_{\rm BA}(\bm k)   &  \hat G_{\rm BB}(\bm k) 
\end{array}
\right).
\end{align}
In particular, the resolvent matrix in the subspace $S_{\rm A}$ can be written as
\begin{equation}
\hat G_{\rm AA}(\bm k,\epsilon)=\left(\epsilon-\hat H^{\rm (eff)}(\bm k,\epsilon)\right)^{-1}
\end{equation}
with the energy dependent effective Hamiltonian
\begin{equation}
\hat H^{\rm (eff)}(\bm k,\epsilon)=\hat H_{\rm AA}+\hat H_{\rm AB}\left(\epsilon-\hat H_{\rm BB}\right)^{-1}\hat H_{\rm BA}.
\label{eq:renorm}
\end{equation}
The poles of $\hat G_{\rm AA}(\bm k,\epsilon)$ gives all the exact eigenenergies of $\hat H(\bm k)$, as long as the projection onto $S_{\rm A}$ of their eigenstates do not vanish.
\par
As in Eq.~(\ref{eq:ham3}), the effective Hamiltonian can be expanded as 
\begin{equation}
\hat H^{\rm (eff)}(\bm k,\epsilon)=E_0(\bm k,\epsilon)\hat I+{\bm R}(\bm k,\epsilon)\cdot\hat{\bm\sigma}^{(m)},
\label{eq:expansion}
\end{equation}
where $E_0$ and ${\bm R}$ are real.
Applying the previous argument to this effective Hamiltonian, we can immediately see that a contact appears at $\bm k=\bm k_0$ and $\epsilon=\epsilon_0$, which is the solution of
\begin{equation}
\epsilon=E_0(\bm k,\epsilon)
\label{eq:cond-1}
\end{equation}
and
\begin{equation}
{\bm R}(\bm k,\epsilon)=\bm 0.
\label{eq:cond-2}
\end{equation}
Equation~(\ref{eq:cond-1}) can be written explicitly as
\begin{align}
&m\epsilon-{\rm Tr}\hat H^{\rm (eff)}(\bm k,\epsilon)\notag\\
&\ \ =m\epsilon-{\rm Tr}\hat H_{\rm AA}-\sum_b w_b(\epsilon-\zeta_b)^{-1}=0,
\label{eq:cond-1a}
\end{align}
with $w_b=\langle b|\hat H_{\rm BA}\hat H_{\rm AB}|b\rangle\ge 0$, where $\zeta_b$ and $|b\rangle$ ($b=1,2,\cdots,n_{\rm s}-2$) denote the eigenenergies and eigenstates of $\hat H_{\rm BB}$, respectively.
Solving this equation with respect to $\epsilon$, we generally find $(n_{\rm s}-m+1)$ real solutions,
\begin{equation}
\epsilon=\xi_j(\bm k)\ \ \ (j=1,2,\cdots,n_{\rm s}-m+1),
\label{eq:sol-1a}
\end{equation}
which satisfy
\begin{equation}
\xi_1<\zeta_1<\xi_2<\zeta_2<\cdots<\zeta_{n_{\rm s}-m}<\xi_{n_{\rm s}-m+1},
\label{eq:sols}
\end{equation}
because the last term in the left-hand side of Eq.~(\ref{eq:cond-1a}) diverges to $\pm\infty$ at $\epsilon\rightarrow\zeta_b\mp 0$ and $\epsilon\rightarrow\pm\infty$.
If some of the eigenenergies of $\hat H_{\rm BB}$ are degenerate, 
the solution exists at that degenerate value, $\xi_{b+1}=\zeta_{b}=\zeta_{b+1}$. 
Strictly speaking, the number of solutions is reduced if there are some $w_b=0$, namely the subspace $S_{\rm A}$ is disconnected from state $|b\rangle$. 
In such case, $|b\rangle$ is the exact eigenstate of the original Hamiltonian, $\hat H(\bm k)$, and can be dealt separately. 
\par
The $m$-fold contact takes place at $\bm k=\bm k_0$, which is a solution of
\begin{equation}
\bm R_j(\bm k)\equiv {\bm R}(\bm k,\xi_j(\bm k))=\bm 0,
\label{eq:cond-2a}
\end{equation}
and its energy is given as $\epsilon_0=\xi_j(\bm k_0)$.
Again, the number of equations is $m^2-1$: $R_{j,1}=R_{j,2}=\cdots=R_{j,m^2-1}=0$.
\par
First, let us consider the $m$-fold contact located at the general $\bm k$-points, which implies that the there are $n_{\rm u}=d$ unknowns, i.e., $(k_1,\cdots,k_d$). 
We further assume that there exist $n_{\rm c}$ constraints near the $\bm k$-points where the contact occurs.
In the vicinity of $\bm k=\bm k_0$, they should be expressed as
\begin{equation}
\bm s^{(i)}_j\cdot\bm R_j(\bm k)=0\ \ \ (i=1,2,\cdots,n_{\rm c}),
\label{eq:nR2}
\end{equation}
within the linear order of $\bm R_j(\bm k)$, where $\bm s_j^{(i)}$ $(i=1,2,\cdots,n_{\rm c})$ is a linear independent set of $(m^2-1)$-dimensional vectors.
The equation (\ref{eq:cond-2a}) is not overdetermined, if 
\begin{equation*}
n_{\rm d}=n_{\rm u}-m^2+1+n_{\rm c}\ge 0,  
\nonumber
\end{equation*}
introduced in advance in Eq.~(\ref{eq:n_d}), is fulfilled, and then the $m$-fold contact of $n_{\rm d}$-dimension becomes feasible. 
Thus, we call Eq.~(\ref{eq:n_d}) {\it feasibility (generalized von-Neumann-Wigner) condition}. 
\par
Next, let us discuss the $m$-fold contact located at the special $\bm k$-points. 
In this case, the number of unknowns decreases: $n_{\rm u}$ must be redefined from the spatial dimension, $d$, to the manifold dimension of the special $\bm k$-points. 
For example, 2D lattices with a reflection symmetry have special $\bm k$-points invariant under the reflection (on the symmetry axis or on the boundary of the Brillouin zone), which results in $n_{\rm u}=1$.
Instead, the number of constraints often increases at the special $\bm k$-points (See \S~(\ref{sec:symmetry})). 
If the Eq.~(\ref{eq:n_d}) holds after this redefinition of $n_{\rm d}$ and $n_{\rm c}$, the equation (\ref{eq:cond-2a}) is not overdetermined, and the $m$-fold and $n_{\rm d}$-dimensional contact at the special $\bm k$-points becomes feasible. 
\par
We can imagine the cases that there exist the maximum number of constraints ($n_c=(m^2\!-\!1)$) on some special $\bm k$-points.
Then, we always obtain $\bm R=\bm 0$ on those special $\bm k$-points, and the contact is classified into the essential degeneracy. 
Otherwise, the contacts occur at some unknown $\bm k$-points, and it is assigned as an accidental degeneracy. 
\par
%*%*%*%*%*%*%*%*%*%*%*%*%*%*%*%*%*%*%*%*%*%*%*%*
\subsection{Dirac points in 2D multi-band systems}\label{sec:Dirac-2D-multiband}
This subsection is devoted to more detailed consideration on the ordinary point contact ($n_{\rm d}=0$, $m=2$) in 2D lattice with more than two bands.
We assume that it is located at a general $\bm k$-point, $\bm k=\bm k_0$ (i.e., $n_{\rm u}=d=2$).
Then, $R_{j,1}(k_1,k_2)=R_{j,2}(k_1,k_2)=R_{j,3}(k_1,k_2)=0$, is fulfilled at $\bm k=\bm k_0$, and the number of the constraints 
required for the feasible contact is one ($n_{\rm c}=n_{\rm d}-n_{\rm u}+m^2-1=1$).
In the vicinity of $\bm k=\bm k_0$, this constraint should be explicitly written as
\begin{equation}
\bm s_j\cdot\bm R_j(\bm k)=0,
\label{eq:nR3}
\end{equation}
in the linear order of $\bm R_j(\bm k)$, where $\bm s_j$ is a nonzero real 3D vector. 
\par
Now, the problem becomes equivalent to the $n_{\rm s}=2$ one discussed in \S\ref{sec:Dirac-2band}. 
Thus, we can derive some general results immediately. 
Dirac points, around which the energy bands always split linearly, are always stable, at which $\bm X_j=\nabla_{k_1}\bm R_j|_{\bm k=\bm k_0}$ and $\bm Y_j=\nabla_{k_2}\bm R_j|_{\bm k=\bm k_0}$ are linearly independent.
As derived in appendix.~\ref{sec:app1}, they are evaluated within the linear order of $\delta\bm k=\bm k-\bm k_0$ as 
\begin{align}
&\epsilon_\pm(\bm k)=\epsilon_0+D^{-1}\left(\bm A\cdot\delta\bm k\pm\sqrt{(\bm R_j\cdot\bm C)^2+D\bm R_j^2}\right)
\label{eq:eqf1}\\
&A_1=D \nabla_{k_1}\xi_j+\bm X_j\cdot\bm C,\hspace{3mm}A_2=D \nabla_{k_2}\xi_j+\bm Y_j\cdot\bm C,\notag\\
&B=\nabla_\epsilon E_0,\notag\hspace{3mm}\bm C=\nabla_\epsilon\bm R,\hspace{3mm}D=(1-B)^2-\bm C^2\ge 1.\notag
\end{align}
Here, all derivatives are evaluated at $\bm k=\bm k_0$ and $\epsilon=\epsilon_0$, and $\bm R_j(\bm k)$ is expanded as $\bm R_j=\bm X_j\delta k_1+\bm Y_j\delta k_2$. 
The elliptic Dirac cone tilts if $\bm A=(A_1,A_2)\ne\bm 0$.
We do not need explicit functional form of $\xi_j(\bm k)$ for the evaluation of $\bm X_j$, $\bm Y_j$, and $\nabla_{\bm k}\xi_j$. 
In fact, they can be explicitly written as
\begin{align}
&\bm X_j=\nabla_{k_1}\bm R+(1-B)^{-1}\left(\nabla_{k_1}E_0\right)\bm C,\notag\\
&\bm Y_j=\nabla_{k_2}\bm R+(1-B)^{-1}\left(\nabla_{k_2}E_0\right)\bm C,\notag\\
&\nabla_{\bm k}\xi_j=(1-B)^{-1}\nabla_{\bm k}E_0.
\label{eq:derivs}
\end{align}
\par
Conversely, if the point contact at $\bm k=\bm k_0$ is unstable, $\bm X_j$ and $\bm Y_j$ are linearly dependent.
In this case, there exists $\delta\bm q\ne 0$ satisfying $\bm X_j\delta q_1+\bm Y_j\delta q_2=\bm 0$, and the band splitting at $\bm k=\bm k_0+\delta\bm k$ shows quadratic dependence on $\delta\bm k$ parallel to $\delta \bm q$.
The Dirac points are always created or destructed as a merged pair under the variation of the lattice parameters, which keeps the constraint of Eq.~(\ref{eq:nR3}).
Thus, also in the multi-band systems we generally find even number of Dirac points between each pair of the adjacent energy bands. 
Under the time-reversal symmetry, the same argument with the one presented in the last paragraph of \S\ref{sec:Dirac-2band} holds. 
\par
In multi-band systems, one band can have Dirac points with both upper and lower bands, which we note upper- and lower-pair, respectively, since each Dirac point has its own pair. 
If the merging of upper-pair occurs at the same special $\bm k$-point with that of the lower-pair, it can be regarded as a new class of merging of Dirac points. 
Figure~\ref{fig:f1}(b) shows the case where upper-pair starts to merge along $\delta \bm q$, while the middle and lower band starts to touch at the same time. 
The latter touching is induced by the merging of lower-pair along different $\delta \bm q$ from the upper-pair one. 
We show in \S\ref{sec:site-centered} the case where four Dirac points actually merge and form a three-fold point contact with one Dirac cone. 
\par
%*%*%*%*%*%*%*%*%*%*%*%*%*%*%*%*%*%*%*%*%*%*%*%*
\subsection{Band overlap}\label{sec:bandoverlap}
In \S \ref{sec:GvNW-theorem} we have shown that the Dirac points can be understood as stable feasible point contacts also in the multi-band systems.
Now, let us assume that the Dirac points exist between the $l$-th and ($l+1$)-th bands, and the system has a commensurate filling factor, $\nu=2l$.
In order to see the interesting physics particular to the Dirac electron systems, the Fermi level should lie exactly at the Dirac points.
Such a situation is realized, only when the $l$-th and $(l+1)$-th bands have no {\it band overlap}, i.e., no hole or electron pockets.
Otherwise, the Fermi level falls off the Dirac point, as shown in Fig.~\ref{fig:f1}(c), and the low energy excitation is dominated by the carriers in the pockets.
\par
At $n_{\rm s}=2$, the band overlap is absent when the diagonal element, $E_0(\bm k)$, in Eq.~(\ref{eq:ham2}) is the $\bm k$-independent constant, since $\epsilon_\pm(\bm k)=E_0 \pm |\bm R(\bm k)|$. 
This condition is realized in a bipartite lattice, where all the diagonal elements ($E_0$ and $R_3$) are $\bm k$-independent. 
However, in $n_{\rm s}=3$ the band overlap is possible even if all the diagonal elements are $\bm k$-independent, e.g., when there is no direct hopping between sites with the same indices $\mu$.
\par
For $n_{\rm s} \ge 3$, we consider the two-fold contact ($m=2$) and again adopt the effective $2\times 2$-Hamiltonian in Eq.~(\ref{eq:renorm}). 
In analogy to $n_{\rm s}=2$, it is straightforwardly concluded that the band overlap is absent if $E_0(\bm k,\epsilon)=\frac{1}{2}{\rm Tr}\hat H^{\rm (eff)}(\bm k,\epsilon)$ is $\bm k$-independent ($\epsilon$-dependence is allowed), i.e., the solutions of Eq.~(\ref{eq:cond-1a}), $\xi_j\; (j=1,2,\cdots,n_{\rm s}-m+1)$ (Eq.~(\ref{eq:sol-1a})), become $\bm k$-independent constants. 
This can be explained as follows; When the $l$-th energy band, $\epsilon_l(\bm k)$, is equal to $\xi_l$, the $l$-th and $(l+1)$-th bands should have a contact at this $\bm k$-point.
Thus, the upper bound of the $l$-th band is given by $\xi_l$.
Similarly, we can also see that the lower bound of the $(l+1)$-th band is given by $\xi_l$.
These facts clearly show that the band overlap never occurs when $\xi_j$'s are constants.
It should be noted that the $\bm k$-independence of $E_0(\bm k,\epsilon)$ is a sufficient condition, and thus one has a chance to find systems without band overlap, even when $E_0(\bm k,\epsilon)$ is $\bm k$-dependent.
We show in the next section the examples of the geometry of lattices which could avoid band overlap and could afford Dirac points at the Fermi level even without this sufficient condition. 
\par
%*%*%*%*%*%*%*%*%*%*%*%*%*%*%*%*%*%*%*%*%*%*%*%*
\subsection{Symmetries and constraints}\label{sec:symmetry}
In the present subsection, we consider the relation between the symmetries and the number of constraints of Eq.~(\ref{eq:nR2}). 
Since the spin-dependent cases are separately discussed in \S\ref{sec:spin}, we neglect the spin-degrees of freedom here. 
The following formulation is applicable to any type of symmetry present in the system.
\par
Let the Hamiltonian, $\hat H(\bm k)$, be invariant under a symmetry operation at a certain $\bm k$-point unchanged by this operation ({\it invariant $\bm k$-point}). 
The effective Hamiltonian, $\hat H^{\rm (eff)}(\bm k,\epsilon)$, keeps this symmetry, if $S_{\rm A}$ is chosen as an invariant subspace.
Throughout this subsection, we consider only the $2\times 2$ effective Hamiltonian ($m=2$) unless otherwise noted. 
There are two possibilities, according to whether $\hat H^{\rm (eff)}(\bm k,\epsilon)$ is invariant under the similarity transformation by a unitary operator, $\mathcal U$, 
\begin{equation}
\mathcal U\hat H^{\rm (eff)}(\bm k,\epsilon)\mathcal U^{-1}=\hat H^{\rm (eff)}(\bm k,\epsilon),\label{eq:unitary}
\end{equation}
or by an antiunitary operator, $\mathcal A$,
\begin{equation}
\mathcal A\hat H^{\rm (eff)}(\bm k,\epsilon)\mathcal A^{-1}=\hat H^{\rm (eff)}(\bm k,\epsilon).\label{eq:antiunitary}
\end{equation}
\par
Equation (\ref{eq:unitary}) implies the presence of a $2\times 2$ unitary matrix, $\hat U$, which fulfills
\begin{equation}
\hat U\hat H^{\rm (eff)}(\bm k,\epsilon)\hat U^{-1}=\hat H^{\rm (eff)}(\bm k,\epsilon).
\label{eq:unitary2}
\end{equation}
Now, remind that any $2\times 2$ unitary matrices can be expressed in the form,
\begin{equation}
\hat U=e^{i\phi}\left(\omega_0\hat I+i\bm\omega\cdot\hat{\bm\sigma}\right),\label{eq:unitary-mat}
\end{equation}
where $\phi$, $\omega_0$, $\bm\omega=(\omega_1,\omega_2,\omega_3)$ are real, and fulfill $\omega_0^2+|\bm\omega|^2=1$ 
(since any unitary matrix can be written as a product of a phase factor and a SU(2) matrix). 
\par
By inserting Eqs.~(\ref{eq:expansion}) and (\ref{eq:unitary-mat}) to Eq.~(\ref{eq:unitary2}), we obtain the commutation relation,
\begin{align}
0=\left[\hat U,\hat H^{\rm (eff)}\right]=\left[\bm\omega\cdot\hat{\bm\sigma},\bm R\cdot\hat{\bm\sigma}\right]=2i\left(\bm\omega\times\bm R\right)\cdot{\bm\sigma},
\end{align}
which immediately gives 
\begin{equation}
\bm\omega\times\bm R=\bm 0.
\label{eq:omega-r}
\end{equation}
Here, we used the formula,
\begin{equation}
(\bm A\cdot\hat{\bm\sigma})(\bm B\cdot\hat{\bm\sigma})=(\bm A\cdot\bm B)\hat I+i(\bm A\times\bm B)\cdot\hat{\bm\sigma}\label{eq:Pauli-mat-formula}.
\end{equation}
Equation (\ref{eq:omega-r}) indicates that $\bm\omega$ should be either zero or parallel to $\bm R$. 
When $\bm\omega=\bm 0$ (i.e., $\hat U=e^{i\phi}\hat 1$), it is trivial that the symmetry gives no constraint. 
As for $\bm\omega\ne\bm 0$, there should be two linearly independent vectors, $\bm s^{(1)}$ and $\bm s^{(2)}$, perpendicular to $\bm\omega$, which serve as two constraints, 
\begin{equation}
\bm s^{(1)}\cdot\bm R=\bm s^{(2)}\cdot\bm R=0.\label{eq:constraint-unitary}
\end{equation}
at the invariant $\bm k$-points. 
\par
Regarding $\mathcal A$ in Eq.(\ref{eq:antiunitary}), we always find a $2\times 2$ unitary matrix, $\hat U$, which satisfies, 
\begin{equation}
\hat U\hat H^{{\rm (eff)}*}(\bm k,\epsilon)\hat U^{-1}=\hat H^{\rm (eff)}(\bm k,\epsilon),
\label{eq:antiunitary2}
\end{equation}
since any antiunitary operators can be expressed as products of complex conjugate and unitary operations. 
By using the expansion,
\begin{equation}
\hat H^{{\rm (eff)}*}=E_0\hat I-\hat\sigma_2(\bm R\cdot\hat{\bm\sigma})\hat\sigma_2,\label{eq:transpose}
\end{equation}
derived from Eq.~(\ref{eq:expansion}), we can rewrite Eq.~(\ref{eq:antiunitary2}) as an anticommutation relation,
\begin{equation}
\left\{\hat U\hat\sigma_2,\bm R\cdot{\bm\sigma}\right\}=0.
\end{equation}
Then, by inserting Eq.~(\ref{eq:unitary-mat}), and using Eq.~(\ref{eq:Pauli-mat-formula}), we obtain
\begin{align}
0
&=\left\{\left(\omega_0\hat I+i\bm\omega\cdot\hat{\bm\sigma}\right)\hat\sigma_2,\bm R\cdot\hat{\bm\sigma}\right\}\notag\\
&=\left\{i\omega_2\hat I+\bm s\cdot\hat{\bm\sigma},\bm R\cdot\hat{\bm\sigma}\right\}\notag\\
&=2i\omega_2\bm R\cdot\hat{\bm\sigma}+2\left(\bm s\cdot\bm R\right)\hat I
\end{align}
with $\bm s=(\omega_3,\omega_0,-\omega_1)$, which gives 
\begin{equation}
\left\{
\begin{array}{l}
\omega_2\bm R=\bm 0\\
\bm s\cdot\bm R=0.
\end{array}\right.
\end{equation}
Thus, we find a single constraint, $\bm s\cdot\bm R=0$, for $\omega_2=0$, and three constraints, $\bm R=\bm 0$, for $\omega_2\ne 0$ at the invariant $\bm k$-points. 
In general, the explicit matrix representation of $\hat U$ is required to judge which of these two cases is realized.
However, if $\mathcal A$ is expressed as a product of a spatial symmetry operation, $\mathcal S$, and time-reversal, $\mathcal T$, one could know the number of constraints without it.
Since $\mathcal S$ is commutable with $\mathcal T$, we obtain $\mathcal S^2=(\mathcal S\mathcal T)^2$, where the right hand side is represented by the matrix, $\hat U \hat U^*$. 
On the other hand, $\omega_2=0$ is equivalent to $\hat U=\hat U^T$, and thus $\hat U\hat U^*=\hat 1$ due to the unitarity, $\hat U^{-1}=\hat U^\dagger=(\hat U^T)^*$. 
These facts show that $\omega_2=0$ is fulfilled when and only when $\mathcal S^2=1$ holds. (Strictly speaking, $\mathcal S^2=1$ must be fulfilled only within the invariant subspace at an invariant $\bm k$-point.)
\par
The representative example of $\omega_2=0$ is the space-time inversion symmetry, i.e., the invariance under the spatial inversion, $\mathcal I$, after the time-reversal, $\mathcal T$, which satisfy $\mathcal I^2=1$, identically.
Both $\mathcal I$ and $\mathcal T$ give rise to the inversion of $\bm k$-points, $\bm k\leftrightarrow -\bm k$, and thus {\it the general $\bm k$-points are kept unchanged} after the space-time inversion. 
Then, we obtain $\omega_2=0$, and a single constraint, $\bm s\cdot\bm R=0$ at every $\bm k$-point.
The 2D lattices with space-time inversion symmetry afford Dirac points at the general $\bm k$-points, because the feasibility condition for the point contact, $n_{\rm d}=n_{\rm u}-m^2+1+n_{\rm c}$, holds there, for $(n_{\rm d},n_{\rm u},m,n_{\rm c})=(0,2,2,1)$.
\par
The example of $\omega_2\ne 0$ is provided by the invariance under the glide reflection, $\mathcal G$ after the time-reversal, $\mathcal T$.
To be more concrete, consider the 2D lattice which is periodic in the $x$- and $y$-directions by lattice constants, $a$ and $b$, respectively.
If this lattice is invariant under the glide reflection in the $x$-axis, i.e., the translation by $(a/2,0)$ after the reflection across the $x$-axis, the special $\bm k$-points on the Brillouin-zone boundary, $k_x=\pm\pi/a$, are symmetry-invariant.
In fact, the operation, $\mathcal{GT}$, change that $\bm k$-points as $(\pm\pi/a,k_y)\rightarrow(\mp\pi/a,-k_y)\equiv(\pm\pi/a,-k_y)\rightarrow (\pm\pi/a,k_y)$.
At these special $\bm k$-points, $\mathcal G^2=\exp\left(2ik_xa/2\right)=-1$ gives $\omega_2\ne 0$.
Actually, we obtain $\omega_2=1$, i.e., $\hat U=e^{i\phi}\sigma_2$, because $\hat U\hat U^*=-\hat 1$ indicates $\hat U=-(\hat U^{-1})^*=-(\hat U^\dagger)^*=-\hat U^T$.
As a result, we obtain $\bm R=\bm 0$, and a line contact is found at the special $\bm k$-points, $k_x=\pi/a$, which is an essential degeneracy.
\par
Sometimes, more than one invariances of Eqs.~(\ref{eq:unitary}) and/or (\ref{eq:antiunitary}) are present at a certain special $\bm k$-point.
Here, we consider the $\pi$-band of graphene, using a tight-binding model of the isotropic honeycomb lattice.
Let us focus one of the unit cells, and assign the two carbon atoms in it as A and B-sites. 
At the K and K'-points, i.e., at the corners of the hexagonal Brillouin zone, Eq.~(\ref{eq:unitary}) holds for both the reflection across the perpendicular bisector of the A-B bond, and the rotation by $2\pi/3$ around the A-site. 
By using the Bloch basis localized on A- and B-sites given in Ref.~\onlinecite{Lomer1955}, the reflection is expressed by the orthogonal matrix, $\hat U_1=\hat\sigma_1$, since it exchanges A- and B-sites in the unit cell. 
By contrast, the rotation do not include the A-B exchange, and is represented by the diagonal unitary matrix, $\hat U_2={\rm diag}(1,\alpha)=\alpha^{-1}(\cos(2\pi/3)\hat 1+i\sin(2\pi/3)\hat\sigma_3)$, where a phase factor, $\alpha=e^{i2\pi/3}$, appears only for the B-site Bloch basis because the B-site moves away to the other unit cell.
These two symmetries requires $\bm\omega_1\times\bm R=\bm\omega_2\times\bm R=\bm 0$ with $\bm\omega_1=(1,0,0)$ and $\bm\omega_2=(0,0,1)$, and thus $\bm R=\bm 0$.
In this way, two point contacts at K- and K'-points in graphene can be understood as essential degeneracies. 
\par
The above essential degeneracies can be understood as a limiting case of the accidental degeneracies. 
In fact, even when the geometrical anisotropy is introduced, i.e., in the absence of the reflection and the three-fold rotational symmetries\cite{wunsch08}, the honeycomb lattice continues to show point contacts at general $\bm k$-points which are the accidental ones under the space-time inversion symmetry. 
Also, the decorated Honeycomb lattice we discuss shortly in \S\ref{sec:reflection2} shows the accidental point contact, in which the decoration site breaks both the space-time inversion and the rotational symmetries but keeps the reflection symmetry instead. 
Therefore, the rotational symmetry in the isotropic honeycomb lattice only works to make the degeneracy essential.
%, and the point contacts themselves are made feasible by the space-time inversion. 
%*%*%*%*%*%*%*%*%*%*%*%*%*%*%*%*%*%*%*%*%*%*%*%*
\subsection{Spin-dependent cases}\label{sec:spin}
Let us briefly discuss here the cases where the spin degrees of freedom affects the orbital degrees of freedom, e.g., the systems under a commensurate spin-density wave (SDW) formation or with spin-orbit interaction. 
In the former example, we should consider the enlarged unit cell with SDW periodicity (magnetic unit cell). 
The most general form of the Hamiltonian is
\begin{equation}
\hat H(\bm k)=
\left(
\begin{array}{cc}
\hat h_{\uparrow\uparrow}(\bm k)&\hat h_{\uparrow\downarrow}(\bm k)\\
\hat h_{\uparrow\downarrow}^\dagger(\bm k)&\hat h_{\downarrow\downarrow}(\bm k)
\end{array}
\right),
\label{eq:hspin}
\end{equation}
where $\hat h_{\sigma\sigma'}(\bm k,\epsilon)$ $(\sigma,\sigma'=\uparrow,\downarrow)$ are $n_{\rm s}\times n_{\rm s}$ matrices.
The matrix dimension of the Hamiltonian is thus doubled. 
Here, refer to {\it spin-dependent} cases as those with $\hat h_{\uparrow\downarrow}(\bm k) \ne 0$ or $\hat h_{\uparrow\uparrow}(\bm k) \ne \hat h_{\downarrow\downarrow}(\bm k)$. 
\par
Special attention should be paid for the space-time inversion symmetry, because the energy bands show the Kramers degeneracy, which is the two-fold essential degeneracy seen at every $\bm k$-point (see appendix.~\ref{sec:app3}). 
We thus need to explore the four-fold degeneracy ($m=4$) in order to consider the contact between two distinct doubly degenerate energy bands.
In Refs.~\onlinecite{fu-kane07} and \onlinecite{murakami07}, the authors investigated the number of parameters required to describe the Hamiltonian with spin-orbit interactions in the context of topological insulators. 
Here, we reinterpret their argument in our context.
\par
To discuss the four-fold contact, we choose two orbitals for each spin in such a way that $S_{\rm A}$ is an invariant subspace of space-time inversion.
Adopting the method mentioned in \S\ref{sec:GvNW-theorem}, we obtain a $4\times 4$ effective Hamiltonian,
\begin{equation}
\hat H^{\rm (eff)}(\bm k,\epsilon)=
\left(
\begin{array}{cc}
\hat h^{\rm (eff)}_{\uparrow\uparrow}(\bm k,\epsilon)&\hat h^{\rm (eff)}_{\uparrow\downarrow}(\bm k,\epsilon)\\
\hat h^{\rm (eff)\dagger}_{\uparrow\downarrow}(\bm k,\epsilon)&\hat h^{\rm (eff)}_{\downarrow\downarrow}(\bm k,\epsilon)
\end{array}
\right).
\label{eq:heff-spin}
\end{equation}
The time-reversal operation, $\mathcal T$, changes the spin-dependent Hamiltonian as
\begin{equation}
\mathcal T \hat H^{\rm (eff)}(\bm k,\epsilon)\mathcal T^{-1}=
\left(
\begin{array}{cc}
\hat h^{{\rm (eff)}*}_{\downarrow\downarrow}(-\bm k,\epsilon)&-\hat h^{{\rm (eff)}T}_{\uparrow\downarrow}(-\bm k,\epsilon)\\
-\hat h^{\rm (eff)*}_{\uparrow\downarrow}(-\bm k,\epsilon)&\hat h^{{\rm (eff)}*}_{\uparrow\uparrow}(-\bm k,\epsilon)
\end{array}
\right),
\label{eq:hspin}
\end{equation}
while the spatial inversions as
\begin{equation}
\mathcal I \hat H^{\rm (eff)}(\bm k,\epsilon)\mathcal I^{-1}\!=\!\left(\!\!
\begin{array}{cc}
\hat U \hat h^{\rm (eff)}_{\uparrow\uparrow}(-\bm k,\epsilon)\hat U^{-1}&\hat U\hat h^{\rm (eff)}_{\uparrow\downarrow}(-\bm k,\epsilon)\hat U^{-1}\\
\hat U\hat h^{\rm (eff)}_{\uparrow\downarrow}(-\bm k,\epsilon)\hat U^{-1}&\hat U\hat h^{\rm (eff)}_{\downarrow\downarrow}(-\bm k,\epsilon)\hat U^{-1}
\end{array}
\!\!\right), 
\end{equation}
with a unitary matrix $\hat U$ satisfying $\omega_2=0$.
Thus, the the space-time inversion symmetry, $\mathcal I\mathcal T\hat H^{\rm (eff)}(\bm k,\epsilon)\mathcal T^{-1}\mathcal I=\hat H^{\rm (eff)}(\bm k,\epsilon)$, implies
\begin{equation}
\left\{
\begin{array}{l}
\hat h^{\rm (eff)}_{\uparrow\uparrow}(\bm k,\epsilon)=\hat U\hat h^{{\rm (eff)}*}_{\downarrow\downarrow}(\bm k,\epsilon)\hat U^{-1},\\
\hat h^{\rm (eff)}_{\uparrow\downarrow}(\bm k,\epsilon)=-\hat U \hat h^{{\rm (eff)}T}_{\uparrow\downarrow}(\bm k,\epsilon)\hat U^{-1}
\end{array}
\right..
\label{eq:site-centered-spin}
\end{equation}
Two spin-diagonal blocks, $h^{\rm (eff)}_{\uparrow\uparrow}(\bm k,\epsilon)$ and $h^{\rm (eff)}_{\downarrow\downarrow}(\bm k,\epsilon)$, are not independent, and we can expand the Hermite matrix, $\hat h^{\rm (eff)}_{\uparrow\uparrow}(\bm k,\epsilon)$, in the form of Eq.~(\ref{eq:ham2}) with four real functions, $E_0$, $R_1$, $R_2$ and $R_3$.
Further, the spin-offdiagonal block is expressed in the form,
\begin{equation}
\hat h^{\rm (eff)}_{\uparrow\downarrow}(\bm k,\epsilon)=\left(Z_1(\bm k,\epsilon)+iZ_2(\bm k,\epsilon)\right)\bm s\cdot\hat{\bm\sigma}
\end{equation}
with two real functions, $Z_1$ and $Z_2$, and the unit vector, $\bm s=(\omega_3,\omega_0,-\omega_1)$.
Therefore, to find a contact, we only need to search the solution of $R_1=R_2=R_3=Z_1=Z_2=0$.
This means that the number of conditions are reduced from $m^2-1=15$ to $5$, and that the space-time inversion symmetry imposes $15-5=10$ constraints on the Hamiltonian.
To have feasible point contacts at the general $\bm k$-points, we still need extra three constraints in two dimension ($n_{\rm u}=d=2$), because the number of necessitated constraints is given as $n_{\rm c}=m^2-1-n_{\rm u}=13$ by Eq.~(\ref{eq:n_d}).
\par
As pointed out in Ref.~\onlinecite{murakami07}, at the special $\bm k$-points, $\bm k=\bm G/2$, we sometimes have larger number of constraints, where $\hat H^{\rm (eff)}(\bm k,\epsilon)$ can become invariant under the time-reversal and inversion, separately. 
This case is discussed in appendix.~\ref{sec:app2}.
\par
%*%*%*%*%*%*%*%*%*%*%*%*%*%*%*%*%*%*%*%*%*%*%*%*
\begin{figure}[tbp]
\begin{center}
\includegraphics[width=8.5cm]{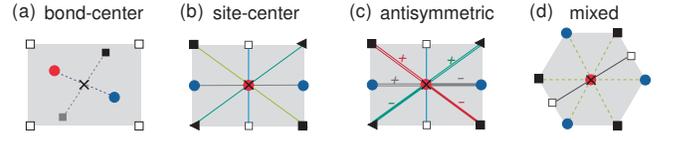}
\end{center}
\caption{(Color online) 
Schematic illustration of (a) bond-centered and (b) site-centered inversions. 
Cross symbols indicate the inversion centers, circles and squares denote the lattice sites, and the shaded regions are the unit cells. 
Pairs of inequivalent sites which exchange by inversion are connected by broken lines. 
Pairs of inequivalent bonds connecting the inequivalent sites which do not exchange by inversion are shown in solid lines. 
Panel (c) is the systems with asymmetric bonds whose Hamiltonian is invariant under the space-time inversion which transforms the phase of site-1 at the inversion center. Panel (d) is the mixture of bond-centered and site-centered sites. 
}
\label{fig:f2}
\end{figure}
%*%*%*%*%*%*%*%*%*%*%*%*%*%*%*%*%*%*%*%*%*%*%*%*
\begin{figure}[tbp]
\begin{center}
\includegraphics[width=8.5cm]{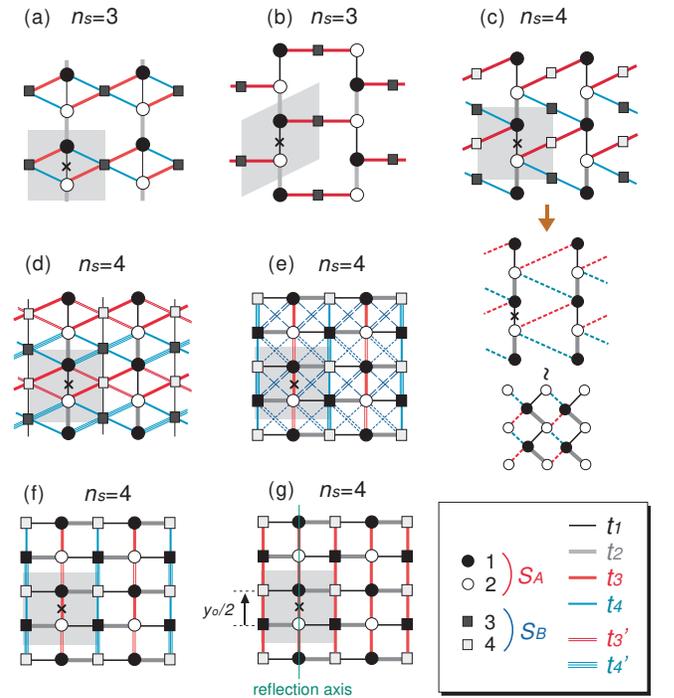}
\end{center}
\caption{(Color online) 
Representative lattice structure with bond-centered inversion of $n_{\rm s}\!=\!3,4$. 
Filled and open circles and squares represent the lattice sites, $\mu=1,2,3,4$, where 1 and 2 belong to subspace $S_{\rm A}$ and others to $S_{\rm B}$. 
Indices of bonds ($t_i$) represented by solid and broken lines are given in the left panel. 
Panel (d) is the model lattice of $\alpha$-ET$_2$I$_3$ which is simplified to panel (c) 
by removing bonds with small transfer integrals. 
The lattice structure of panel (c) is reduced to that of panel (b) when site-4 is removed. 
If we take the vertical bonds in (f) uniform as, $t_3=t_3'=t_4=t_4'$, we find the lattice in (g) which has glide reflection symmetry ($y_o/2$-translation in the vertical direction and reflection with respect to $y$-axis) and affords essential line degeneracy. 
}
\label{fig:f3}
\end{figure}
%*%*%*%*%*%*%*%*%*%*%*%*%*%*%*%*%*%*%*%*%*%*%*%*
\begin{figure}[tbp]
\begin{center}
\includegraphics[width=8.5cm]{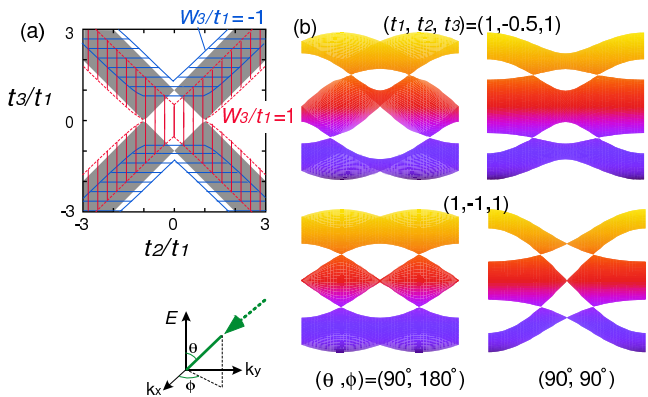}
\end{center}
\caption{(Color online) 
Examination of the point contacts of the lattice structure in Fig.~\ref{fig:f3}(b). 
(a) $t_{2}/t_1$-$t_{3}/t_1$ diagram; the Dirac points are stable in the shaded region at $W_\mu=0, (\mu=1-3)$. 
These stable regions shifts to those indicated by hatches when $W_3/t_1=1$ and -1, respectively. 
(b) Examples of band structures with Dirac points in the shaded region in (a), which are viewed from two different directions, $(\theta,\phi)=(90^\circ,180^\circ)$ and $(90^\circ,90^\circ)$. 
The polar angles, $\theta$ and $\phi$, are defined in ($k_x$,$k_y$,$\epsilon$)-space in the panel, where arrows indicate the direction of a view. 
}
\label{fig:f4}
\end{figure}
%*%*%*%*%*%*%*%*%*%*%*%*%*%*%*%*%*%*%*%*%*%*%*%*
\begin{figure}[tbp]
\begin{center}
\includegraphics[width=8.5cm]{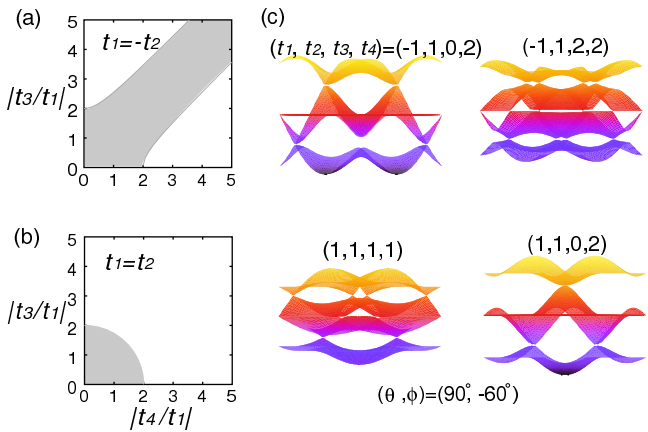}
\end{center}
\caption{(Color online) 
Examination of the point contacts of the simplified $\alpha$-type lattice structure in Fig.~\ref{fig:f3}(d). 
$|t_3/t_1|$-$|t_4/t_1|$ diagram for (a)$t_{1}\!=\!-t_{2}$ and (b)$t_{2}\!=\!t_{1}$. 
Dirac points are stable in the shaded regions ((a): $\sqrt{(t_{4}/t_1)^2-(t_{3}/t_1)^2} \le 2$ and (b): $(t_{3}/t_1)^2+(t_{4}/t_1)^2 \le 4$). 
Panels in (c) show four different examples of the band structures with Dirac points. 
}
\label{fig:f5}
\end{figure}
%*%*%*%*%*%*%*%*%*%*%*%*%*%*%*%*%*%*%*%*%*%*%*%*
%*%*%*%*%*%*%*%*%*%*%*%*%*%*%*%*%*%*%*%*%*%*%*%*
\section{tight-binding models}\label{sec:Dirac}
In this section, we demonstrate that the feasibility (generalized von-Neumann-Wigner) condition, Eq.~(\ref{eq:n_d}), is useful to design Dirac points in unexplored lattices. 
We consider tight-binding model, which includes $n_{\rm s}$ atomic orbitals per unit cell. 
Its Hamiltonian has transfer integrals between the sites and the potentials, $W_\nu$, on the $\nu$-th site (while we set $W_\nu=0$ unless otherwise noted). 
We basically consider single orbital per lattice site, but can deal with the multi-orbital ones, since the sites with $l$ atomic orbitals can be regarded as $l$ inequivalent sites at the same position. 
We introduce Bloch basis, $|\mu,\bm k\rangle$, which has a probability amplitude only at the $\mu$-th site in the unit cell ($\mu,\nu=1,2,\cdots, n_{\rm s}$). 
Then, the matrix elements of Hamiltonian is given as $\hat H_{\mu\nu}(\bm k)=\langle \mu,\bm k | {\mathcal H} |\nu,\bm k\rangle$. 
\par
The practical procedure to explore the $\bm k$-points is given as follows. 
Suppose we examine whether a given $d$-dimensional lattice structure could afford $m$-fold and $n_{\rm d}$-dimensional contact. 
Here, the lattice is characterized by a set of lattice parameter, $\bm t$, which is dealt explicitly as variables. 
\vspace{-2mm}
\begin{itemize}
\item[(i)] Count the number of constraints, $n_{\rm c}$, on the general $\bm k$-point ($n_{\rm u}=d$). 
(If the special $\bm k$-points have larger number of constraints, separately count $n_{\rm c}$ there, and redefine $n_{\rm u}$ as the dimension of manifold formed by these special $\bm k$-points). 
\vspace{-2mm}
\item[(ii)]Check whether the feasibility condition Eq.(\ref{eq:n_d}) is fulfilled. 
\end{itemize}
\vspace{-2mm}
If fulfilled, we go onto the next step to specify the region of lattice parameter space with stable contacts, and to know the location of contacts in $\bm k$-space, $\bm k=\bm k_0$. 
\vspace{-2mm}
\begin{itemize}
\item[(iii)]
Start from the $n_{\rm s}\times n_{\rm s}$ Hamiltonian, which is an explicit function of $\bm t$ besides $\bm k$. 
Derive an effective $m\times m$ Hamiltonian, $\hat H^{\rm (eff)}(\bm k,\epsilon,\bm t)$, using Eq.(\ref{eq:renorm}). 
\vspace{-2mm}
\item[(iv)]
Eqs.(\ref{eq:cond-1}) and (\ref{eq:cond-2}) give the {\it analytical relationship} between $\bm t$ and $\bm k_0$, when the contacts are present. 
\end{itemize}
\vspace{-2mm}
One can thus identify the region with contacts within the lattice parameter space. 
Inside this identified region, the contacts are stable, whereas on its edge the contact becomes unstable. 
The examples given in this section follow these treatments. 
\par
In the 2D lattices shown in Figs.\ref{fig:f3}, \ref{fig:f6}, and \ref{fig:f8}, the site indices $\mu=1,2,3,4$ are assigned to those represented by filled circles, open circles, filled squares, and open squares, respectively. 
For the two-fold degeneracies ($m=2$), we consider the 2D subspace, $S_{\rm A}$, spanned by $\mu=1,2$. 
The $m=3$ fold contact is also discussed for the Lieb lattice. 
\S\ref{sec:inversion} and \S\ref{sec:reflection2} are devoted to the spin-independent systems, 
and the spin-dependent example is presented in \S\ref{sec:spin}. 
\par
%*%*%*%*%*%*%*%*%*%*%*%*%*%*%*%*%*%*%*%*%*%*%*%*
\subsection{Space-time inversion}\label{sec:inversion}
In the present subsection, we specify the discussions on the space-time inversion symmetry to the tight-binding models, 
where the types of inversion symmetry are classified by their appearance of lattices which directly reflect the shapes and locations of atoms in the crystal. 
We assume without the loss of generality that the atomic orbitals have either odd or even parities under the inversion operation. 
Let us regard a pair of sites as equivalent or inequivalent, according to whether or not they are different only by a lattice vector. 
Inversion invariance of the lattice are basically classified into two cases: the {\it bond-} and {\it site-centered} inversions, which are shown schematically in Figs. \ref{fig:f2}(a) and \ref{fig:f2}(b), respectively. 
If the inversion symmetry is accompanied by the exchange of the pair of inequivalent sites, it is called {\it bond-centered} one. 
In fact, the inversion center should locate at the middle point of the bond which connects that pair. 
On the other hand, the inversion symmetry without exchange of inequivalent sites is called {\it site-centered} one, since the inversion center can be placed at one of these sites. 
There is one more class of lattices which differ from the previous ones by its appearance, shown in Fig.\ref{fig:f2}(c), 
where the bonds exchange their sign (but not their absolute values) by the inversion operation. 
If the system includes pairs of inequivalent sites with and without the exchange (see e.g., Fig.\ref{fig:f2}(d)), one can separately choose each pair as the Hilbert subspace $S_{\rm A}$, to which the bond-centered and site-centered inversion symmetry are adopted, respectively. 
Both cases will afford feasible contacts. 
\par
In the bond-centered inversion, the Hilbert subspace $S_{\rm A}$ is spanned by $|1,\bm k\rangle$ and $|2,\bm k\rangle$, where $\mu=1$ and $2$ are the indices for the pair of the corresponding atomic orbitals on the inequivalent sites exchanged by the inversion operation.
In this case, the space-time inversion symmetry requires
\begin{equation}
\hat H^{\rm (eff)}=\mathcal I\mathcal T\hat H^{{\rm (eff)}}\mathcal T^{-1}\mathcal I^{-1}=\hat\sigma_1\hat H^{{\rm (eff)}*}\hat\sigma_1,
\label{eq:bond-inv}
\end{equation}
leading to $(\omega_0,\bm\omega)=(0,1,0,0)$, where $\hat\sigma_1$ denotes the exchange between two inequivalent sites.
Thus, we immediately find a constraint, $\bm s\cdot\bm R_j=0$, with $\bm s=(0,0,1)$ at general $\bm k$-points.
\par
In the site-centered inversion, on the other hand, the subspace $S_{\rm A}$ is spanned by $|1,\bm k\rangle$ and $|2,\bm k\rangle$, where $\mu=1$ and 2 are the indices of an arbitrary pair of inequivalent atomic orbitals which remain unchanged by inversion.
In this case, we can see 
\begin{equation}
\hat H^{\rm (eff)}=\mathcal I\mathcal T\hat H^{{\rm (eff)}}\mathcal T^{-1}\mathcal I^{-1}=\hat H^{{\rm (eff)}*},
\label{eq:site-inv}
\end{equation}
with $(\omega_0,\bm\omega)=(1,0,0,0)$.
Thus, we have a constraint, $\bm s\cdot\bm R_j=0$, with $\bm s=(0,1,0)$ at general $\bm k$-points.
\par
The remaining issue is the inversion represented by Fig.\ref{fig:f2}(c). 
The subspace $S_{\rm A}$ is spanned by $|1,\bm k\rangle$ and $|2,\bm k\rangle$, where $|1,\bm k\rangle$ is connected to $|\nu,\bm k\rangle$ ($\nu=2\sim n_s$) by the pairs of bonds which exchange their sign by inversion. 
Namely, the inversion operates in such a way that the phase of $|1,\bm k\rangle$ is shifted by $\pi$ as, 
\begin{equation}
\hat H^{\rm (eff)}=\mathcal I\mathcal T\hat H^{{\rm (eff)}}\mathcal T^{-1}\mathcal I^{-1}=\hat\sigma_3\hat H^{{\rm (eff)}*}\hat\sigma_3. 
\label{eq:bond-sign-ex}
\end{equation}
Such case can be realized when $|1,\bm k\rangle$ and $|2,\bm k\rangle$ are spatially anisotropic and isotropic atomic orbitals (e.g. $d$- and $s$-orbitals), respectively. 
%*%*%*%*%*%*%*%*%*%*%*%*%*%*%*%*%*%*%*%*%*%*%*%*
\subsubsection{Bond-centered inversion}\label{sec:bond-centered}
Typical lattice structures with the bond-centered inversions are given in Figs.~\ref{fig:f3}(a)-\ref{fig:f3}(f). 
All these lattices fulfill the feasibility condition and could afford Dirac points in the certain range of lattice parameter space. 
\par
Let us first examine the staggered square lattice in Fig.~\ref{fig:f3}(b). 
We find that the Dirac points appear in the shaded region displayed in Fig.~\ref{fig:f4}(a) on the plane of $t_{2}/t_{1}$ and $t_{3}/t_{1}$ at $W_\mu/t_1=0$. 
The on-site potential, $W_3$, which does not break the inversion, keeps the region with stable Dirac points, 
as the figure shows for the cases with $W_3/t_1=\pm 1$. 
The Dirac points in this region are indeed stable against the variation of lattice parameters including the on-site potentials. 
\par
Next, we add one extra site to Fig.~\ref{fig:f3}(b), which yields a lattice shown in Fig.~\ref{fig:f3}(c). 
This lattice is a simplification of $\alpha$-ET$_2$I$_3$ shown in Fig.~\ref{fig:f3}(d). 
The parameter regions which afford Dirac points in Fig.~\ref{fig:f3}(c) are examined 
for $t_{1}=-t_{2}$ and $t_{1}=t_{2}$ in Figs.~\ref{fig:f5}(a) and \ref{fig:f5}(b), respectively. 
The region with Dirac points of the former is extensive over the wide parameter space, 
$\sqrt{(t_4/t_1)^2-(t_3/t_1)^2} \le 2$, 
which spreads along $0<|t_3/t_1|\sim |t_4/t_1|<+\infty$. 
Whereas, the region in the latter is confined to, $(t_3/t_1)^2+(t_4/t_1)^2 < 4 $, 
namely within the small value of $|t_3/t_1|,\:|t_4/t_1| \le 2$. 
It should be noted that parameter values of $\alpha$-ET$_2$I$_3$ corresponds to the former case\cite{alphaI3}. 
Figure~\ref{fig:f5}(c) shows the band structures for four different choices of lattice parameters. 
The last panel is the one on the edge of the Dirac point region in Fig.~\ref{fig:f5}(b) (unstable Dirac point), 
where the two Dirac points merge and the dispersion in the merging direction become parabolic. 
\par
Among the lattices with bond-centered inversions, the ones in Fig.~\ref{fig:f3}(b), \ref{fig:f3}(c), and \ref{fig:f3}(f) have $H^{\rm (eff)}(\bm k, \epsilon)$ with $\bm k$-independent diagonal elements and thus have no band overlap. 
One can generalize these cases and refer to it as, {\it ``the effective bipartite lattices have no band overlap''}, where the words, ``effective bipartite lattices'', mean that the lattices have no direct hopping between site-1 and -2, and at the same time, no indirect path, $1\rightarrow \nu \rightarrow 1$ or $2\rightarrow \nu \rightarrow 2$  ($\nu=3,4$). 
In such case, the diagonal elements of $H_{\rm AA}$ has no $\bm k$-dependence and the second term of Eq.~(\ref{eq:renorm}) also has no diagonal matrix elements. Thus, the effective Hamiltonian indeed fulfills the sufficient condition to avoid band overlap. 
\par
In the lower panel of Fig.~\ref{fig:f3}(c), the sites $\nu=3,4$ and 
their related bonds 1$\rightarrow \nu \rightarrow$1 and 2$\rightarrow \nu \rightarrow$2 
are replaced by the the dotted lines, which schematically describes the effective bipartite lattice. 
This gives the physical interpretation of Eq.~(\ref{eq:renorm});
Its second term formally describes the hopping of particle between site-1 and -2 in $S_{\rm A}$ 
mediated by the occupation of particles at site-3 and -4 in $S_{\rm B}$, 
and is regarded as having effective transfer integrals between 1$\rightarrow$1 and 2$\rightarrow$2. 
The resultant effective lattice, simplified from the $\alpha$-type one, 
is topologically equivalent to the anisotropic square lattice in the lowest panel, 
which is known to afford Dirac points in a wide parameter region, 
as discussed in the context of $\alpha$-ET$_2$I$_3$\cite{katayama06}. 
\par
Finally, we briefly discuss the case with more than one invariances. 
Figure~\ref{fig:f3}(f) has a bond-centered inversion and affords feasible contact points. If all the vertical bonds are taken uniform as $t_3=t_3'=t_4=t_4'$, one finds Fig.~\ref{fig:f3}(g) which has the glide reflection symmetry (see \S\ref{sec:symmetry} for details) in addition to inversion. 
By the translation of half the unit cell length in the vertical direction together with the reflection against the vertical axis, 
the site exchanges as $1 \leftrightarrow 2$ and $3 \leftrightarrow 4$, while the bonds remain unchanged. 
The special $\bm k$-points invariant under this glide reflection is $k_x=\pi$. 
In fact, one can easily check that $H^{{\rm (eff)}}$ spanned by $S_{\rm A}=\{1,2\}$ automatically fulfills Eqs.(\ref{eq:cond-1}) and (\ref{eq:cond-2}) at $k_x=\pi$, which indicates the realization of essential degeneracy. 
The essential degeneracy thus can often be understood as a limiting cases of feasible accidental degeneracy under the variation of lattice parameters due to the introduction of additional invariance, which we have also seen in the honeycomb lattice of graphene in \S\ref{sec:symmetry}. 
\par
%*%*%*%*%*%*%*%*%*%*%*%*%*%*%*%*%*%*%*%*%*%*%*%*
\begin{figure}[tbp]
\begin{center}
\includegraphics[width=8.5cm]{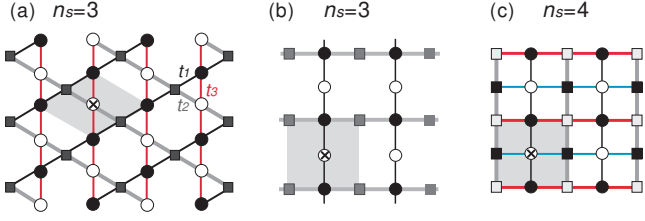}
\end{center}
\caption{(Color online) 
Representative lattice structure with site-centered inversion of $n_{\rm s}\!=\!3,4$, 
(a) Kagom\'e, (b) Lieb, and (c) anisotropic square lattices. 
The notation of lattice sites and the transfer integrals follow those given in Fig.~\ref{fig:f3}.
}
\label{fig:f6}
\end{figure}
%*%*%*%*%*%*%*%*%*%*%*%*%*%*%*%*%*%*%*%*%*%*%*%*
\begin{figure}[tbp]
\begin{center}
\includegraphics[width=8.5cm]{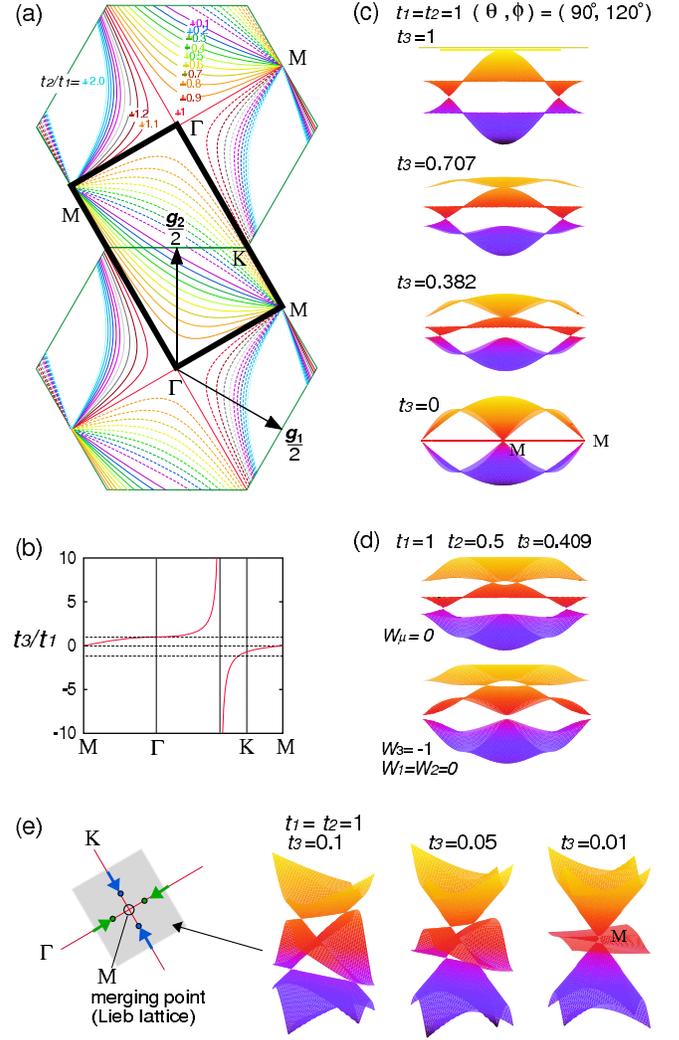}
\end{center}
\caption{(Color online) 
Dirac points of the Kagom\'e lattice, 
which merge at $\Gamma$ and M-points, but do not open a gap throughout the whole $(t_1,t_2,t_3)$-space, and under the on-site potentials. 
(a) Trajectories of point contacts between upper two bands in the anisotropic Kagom\'e lattice in Fig.~\ref{fig:f6}(a) under the variation of $t_3/t_1$ for several choices of $0\le t_2/t_1\le 2$. 
(b) Variation of $t_3/t_1$ along the M-$\Gamma$-K-M points which are indicated in bold lines in panel (a). 
(c) Variation of band structures along the $\Gamma$-M line with $t_2/t_1=1$. The last panel is the energy bands of the Lieb lattice. 
(d) Examples of band structure with and without on-site potential. 
(e) Merging of four Dirac points in the vicinity of M-point under $t_2/t_1=1$, $t_3/t_1\rightarrow 0$. Their cross sections along the different merging directions of upper- and lower-pairs correspond to those in Fig.~\ref{fig:f1}(b).
}
\label{fig:f7}
\end{figure} 
%*%*%*%*%*%*%*%*%*%*%*%*%*%*%*%*%*%*%*%*%*%*%*%*
\begin{figure}[tbp]
\begin{center}
\includegraphics[width=8.5cm]{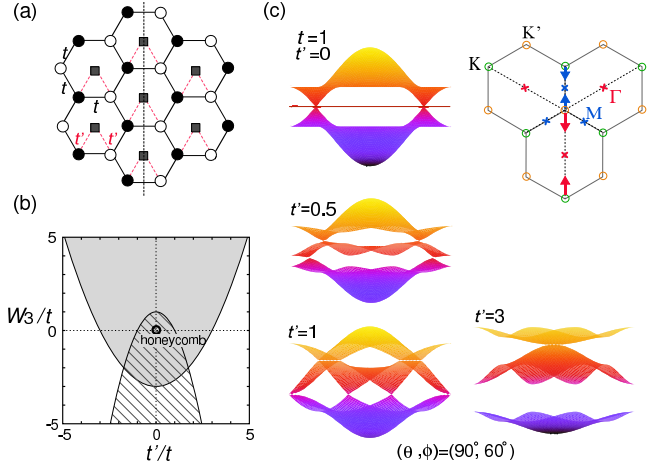}
\end{center}
\caption{(Color online) 
(a) Decorated honeycomb lattice structure of $n_{\rm s}=3$ without the inversion symmetry but with reflection symmetry. 
(b) Upper shaded ($W_3/t\ge (t'/t)^2/3-3$) and lower hatched ($W_3/t\le -(t'/t)^2+1$) regions afford Dirac points between the upper two and lower two bands, respectively, and the region (shaded + hatched) at the center have the two sets of Dirac point. 
(c) Demonstration of the existence of Dirac points under the reflection symmetry, which emerge along the K-K'-$\Gamma$ lines in $\bm k$-space. 
The contacts on K- and K'-points at $t'=0$(isotropic honeycomb lattice) and those at $t'/t\ne 0$ are essential and accidental ones, respectively. 
The upper and lower pairs merge at $\Gamma$- and $M$-points, respectively, at $t'/t=\pm 3$ and $\pm 1$. 
}
\label{fig:f8}
\end{figure} 
%*%*%*%*%*%*%*%*%*%*%*%*%*%*%*%*%*%*%*%*%*%*%*%*
%*%*%*%*%*%*%*%*%*%*%*%*%*%*%*%*%*%*%*%*%*%*%*%*
\subsubsection{Site-centered inversion}\label{sec:site-centered}
The representative lattice structures with the site-centered inversion symmetry are the Kagom\'e, Lieb, and the anisotropic square lattices shown in Figs.~\ref{fig:f6}(a)-\ref{fig:f6}(c). 
Noteworthy is the Dirac points in the Kagom\'e lattice, which is found {\it over the whole parameter region of the anisotropy of transfer integrals and on-site potentials}. 
Under the variation of the transfer integrals, the Dirac points appear over the whole $\bm k$-space. 
Figure~\ref{fig:f7}(a) shows the trajectories of Dirac $\bm k$-points between upper two bands 
for the several fixed values of $0\le t_{2}/t_3 \le 2$, under the variation of $t_{1}/t_3$. 
The values of $t_{1}/t_3$ to have Dirac points along the M-$\Gamma$-X-M line are shown in Fig.~\ref{fig:f7}(b). 
In the Kagom\'e lattice, there always exist two sets of Dirac points between upper two and lower two bands 
at $\bm k=\pm \bm k_0^+$ and $\pm \bm k_0^-$, respectively.  
Here, these $\bm k$-points are described in the form, $\bm k_0^+=(k_1^0\bm g_1+ k_2^0\bm g_2)/2$, and $\bm k_0^-=(k_1^0\bm g_1 - k_2^0\bm g_2)/2$, with the reciprocal lattice vectors, $\bm g_1$ and $\bm g_2$ (see Fig.\ref{fig:f7}(a)). 
Figures ~\ref{fig:f7}(c) and \ref{fig:f7}(d) show the energy bands for several choices of lattice parameters. 
In the regular Kagom\'e lattice with isotropic transfer integrals (see the first panel of Fig.~\ref{fig:f7}(c)), 
one pair of Dirac points between the upper two bands merge at $\Gamma$-point and touch the flat band in the parabolic manner. 
\par
When one of the three bonds in the Kagom\'e lattice is subtracted (e.g., $t_{3}\rightarrow0$), the Lieb lattice is realized. 
In this limit, two Dirac points between the upper two bands and another two between the lower two bands merge at M-point ($\bm k_0=(\bm g_1 + \bm g_2)/2$), and a three-fold contact appears. 
Around the three-fold point contact, the center band is flat, and the upper- and lower- Dirac cones touch it at their tips.
The merging of four Dirac points is shown in Fig.~\ref{fig:f7}(e) in the vicinity of M-point. 
This particular class of band touching takes place regardless of the values of $t_1,t_2(\ne 0)$. 
\par
It is interesting to confirm that the feasibility condition, $n_{\rm d}=n_{\rm u}-m^2+n_{\rm c}=0$, in Eq.~(\ref{eq:n_d}) holds for the three-fold point contact at general $\bm k$-point ($n_{\rm d}=0$, $n_{\rm u}=2$ and $m=3$) in the Lieb lattice.
The Hamiltonian, $\hat H(\bm k)$, is a $3\times 3$ Hermite matrix with only two nonzero elements, $H_{12},H_{13}$, which can be expanded as Eq.~(\ref{eq:ham3}) with Gell-Mann matrices, Eq. (\ref{eq:gellmann}).
We immediately see $R_3=R_8=0$ from $H_{\mu\mu}=0$, and $R_6=R_7=0$ from $H_{23}=0$.
Further, the space-time inversion symmetry offers $R_2=R_5=R_7=0$. 
As a result, one finds $n_{\rm c}=6$ in all, and sees that the feasibility condition is fulfilled: $n_{\rm d}=2-3^2+1+6=0$.
Although the Lieb lattice can have the three-fold point contact at a general $\bm k$-point, it always locates at M-point, a special $\bm k$-point, 
which is because it is generated by the merging of Dirac points. 
\par
We finally mention that the anisotropic square lattice in Fig.~\ref{fig:f6}(c) is {\it effectively bipartite}, and thus has not band overlap. 
The Kagom\'e and Lieb lattices also do not have band overlaps, even though the trace of $\hat H^{\rm (eff)}(\bm k,\epsilon)$ have $\bm k$-dependence 
and do not fulfill the sufficient condition to avoid band overlap. 
\par
%*%*%*%*%*%*%*%*%*%*%*%*%*%*%*%*%*%*%*%*%*%*%*%*
\subsection{Reflection}\label{sec:reflection2}
The final non-interacting example is devoted to the decorated honeycomb lattice shown in Fig.~\ref{fig:f8}(a). 
Here, the inversion symmetry is broken by the decoration, namely an introduction of the third site described by square symbol, whereas, the reflection symmetry is retained, which exchanges $\mu=1,2$. 
\par
In contrast to the space-time inversion, our consideration should be restricted to the {\it special $\bm k$-points on the symmetry axis}, which is unchanged by the reflection.
It requires
\begin{equation}
\hat H^{\rm (eff)}=\mathcal R\hat H^{\rm (eff)}\mathcal R^{-1}=\hat\sigma_1\hat H^{\rm (eff)}\hat\sigma_1,
\label{eq:reflection}
\end{equation}
giving $(\omega_0,\bm\omega)=(0,1,0,0)$, where the subspace $S_{\rm A}$ is spanned by $|1,\bm k\rangle$ and $|2,\bm k\rangle$. 
Thus, we can find two constraints, $\bm s^{(1)}\cdot\bm R_j=\bm s^{(2)}\cdot\bm R_j=0$, with $\bm s^{(1)}=(0,1,0)$ and $\bm s^{(2)}=(0,0,1)$ on the special $\bm k$-points invariant under the reflection, located on the symmetry axis or on the boundary of Brillouin zone.
\par
We expect to find a feasible point contact on the special $\bm k$-points of symmetry axis, because the number of unknowns and constraints are $n_{\rm u}=1$, and $n_{\rm c}=2$, respectively, giving $n_{\rm d}=0$ in Eq.~(\ref{eq:n_d}).
Figure~\ref{fig:f8}(b) shows the parameter region with Dirac points on the plane of $t'/t$ and $W_{3}/t$, whose origin correspond to the regular (isotropic) honeycomb lattice. 
The shaded and hatched regions afford Dirac points between the upper two and lower two bands, respectively. 
In Fig.~\ref{fig:f8}(c), we show several examples of Dirac points at $W_3=0$: 
When $t'/t=0$, the {\it essential} Dirac points between the top and bottom bands are located at K- and K'-points at the corner of the hexagonal Brillouin zone. 
By the introduction of $|t'/t|>0$, these Dirac points split into two pairs and become {\it accidental}. With increasing $|t'/t|$ they move along the symmetric axis connecting K-, K'-, and $\Gamma$-points. The lower pair merge first at M-point (unstable point at $|t'/t|=1$) and dissapear. Then, the upper pair merge at $\Gamma$ point ($|t'/t|=3$) and dissapear. 
\par
%*%*%*%*%*%*%*%*%*%*%*%*%*%*%*%*%*%*%*%*%*%*%*%*
\subsection{Dirac points in SDW}\label{sec:SDW}
We discuss the spin-density-wave (SDW) state as an example of the spin-dependent system, with LaOFeAs in mind. 
This material has five orbitals per site\cite{kuroki08} and exhibits Dirac points in the vicinity of the Fermi level. 
Actually, they are reproduced within the self-consistent solution of the mean-field approximation of the five-orbital Hubbard model with on-site interactions, $U$, $U'$, and $J$, which denotes the intra-orbital-direct, inter-orbital-direct, inter-orbital-exchange Coulomb interactions, respectively\cite{kuroki08,ran09}.
\par
\begin{figure}[tbp]
\begin{center}
\includegraphics[width=8.5cm]{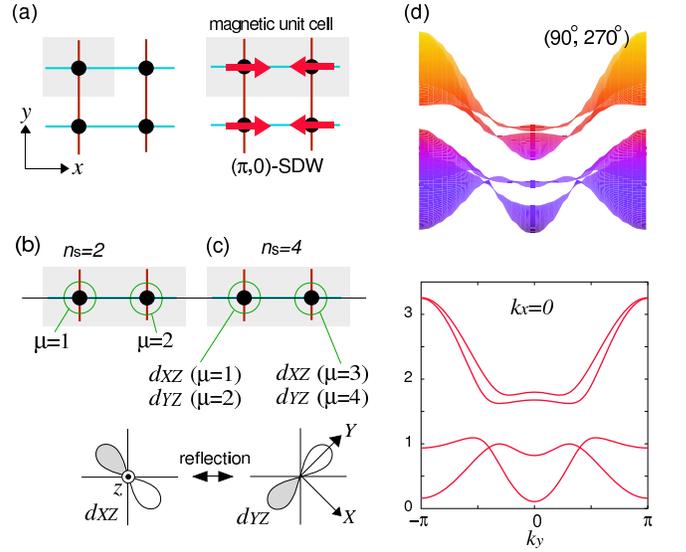}
\end{center}
\caption{(Color online) 
(a) Square lattice with original unit cell and magnetic unit cell of SDW state. Arrows represent spins which have $(\pi,0)$-periodicity. 
Magnetic unit cell with (b) $n_{\rm s}$=2 and (c) $n_{\rm s}=4$. 
For the case with (c), Hamiltonian is invariant under the reflection against either the $xz$- or $yz$-plane, both of which exchange the two orbitals, $d_{XZ}$ and $d_{YZ}$. 
(d) Energy band of the self-consistent SDW solution for parameters, 
$t_{\nu\mu}=-0.2$(nearest neighbor sites), $(t'_{13},t'_{24})=(0.3,0.15)$ (next-nearest neighbor sites in the (1,1)-direction), $U=1.2$, $U'=0.9$, and $J=0.15$. 
The left panel is the cross-section at $k_x=0$ as a function of $k_y$, along which the Dirac point emerges (symmetry axis). 
}
\label{fig:f9}
\end{figure}

Here, we consider the simpler two-orbital Hubbard model on an isotropic square lattice, as shown in Fig.~\ref{fig:f9}(a), and focus only on the $(\pi,0)$-SDW, which has two-fold periodic spin modulation along the $x$-direction\cite{ran09}.
The magnetic unit cell is doubled from the original one, and includes four orbitals.
The spin-dependent mean-field Hamiltonian is an $8\times 8$ matrix, and can be reduced to the $4\times 4$ effective one described by Eq.~(\ref{eq:hspin}).
The space-time symmetry gives 10 constraints on the effective Hamiltonian as discussed in \S\ref{sec:spin}.
Further, there are two additional constraints, $Z_1=Z_2=0$, because the interaction processes do not mix states with opposite spins ($\hat h^{\rm (eff)}_{\uparrow\downarrow}=0$).
As a result, we can focus only on the $2\times 2$ matrix, $h^{\rm (eff)}_{\uparrow\uparrow}$, to search for the contact, and the problem becomes equivalent to the spin-independent case. 
The only difference is that the space-time inversion symmetry is already used to reduce the problem, and thus imposes no constraint on $h^{\rm (eff)}_{\uparrow\uparrow}$.
\par
It is noteworthy that we need more than one orbital per site in order to find Dirac points in the SDW state.
To show this point, let us consider a single orbital Hubbard model (see Fig.~\ref{fig:f9}(b)), where $h_{\uparrow\uparrow}$ in the original mean-field Hamiltonian already has a $2\times 2$ matrix form in Eq.~(\ref{eq:ham2}).
Since inter-orbital interactions are absent ($U'=J=0$), the on-site interaction is simply evaluated as $Un_{\mu\uparrow}n_{\mu\downarrow}\sim U\left(\langle n_{\mu\uparrow}\rangle n_{\mu\downarrow}+n_{\mu\uparrow}\langle n_{\mu\downarrow}\rangle-\langle n_{\mu\uparrow}\rangle\langle n_{\mu\downarrow}\rangle\right)$, where $\mu=1,2$ denotes the index for the atomic orbitals in the magnetic unit cell, and $\sigma=\uparrow,\downarrow$ denotes the electron spin.
This results in $R_3=U\left(\langle n_{1\downarrow}\rangle-\langle n_{2\downarrow}\rangle\right)/2 \ne 0$ in the expansion of $h_{\uparrow\uparrow}=E_0\hat I+\bm R\cdot\bm {\hat\sigma}$, and implies that the contact, i.e., the solution of $\bm R=\bm 0$, is never found at any $\bm k$-point.
\par
Let us return to our main point, and consider the two-orbital Hubbard model, where the two atomic orbitals have $d_{XZ}$- and $d_{YZ}$-symmetries as shown in Fig.~\ref{fig:f9}(c).
This model is invariant under the reflection across $x$- or $y$-axis.
We generate an effective $2\times 2$ Hamiltonian, $h^{\rm (eff)}_{\uparrow\uparrow}(\bm k,\epsilon)$, focusing on two atomic orbitals ($m=2$) in the original unit cell ($\mu=1,2$).
Since the reflection exchanges these atomic orbitals, $1\leftrightarrow 2$, $h^{\rm (eff)}_{\uparrow\uparrow}=\hat\sigma_1 h^{\rm (eff)}_{\uparrow\uparrow}\hat\sigma_1$ holds on the symmetry axis ($n_{\rm u}=1$), which leads to two constraints ($n_{\rm c}=2$), $R_2=R_3=0$.
This situation is equivalent to the case discussed in \S\ref{sec:reflection2}, and thus the feasibility condition, Eq.~(\ref{eq:n_d}), is fulfilled on the symmetry axis.
Figure ~\ref{fig:f9}(d) shows one of the examples of self-consistent solutions of SDW by choosing the parameter close to those given in Ref.~\onlinecite{kuroki08}. 
The lowest two bands do show point contacts located on the symmetry axis ($k_x=0$). 
\par
%*%*%*%*%*%*%*%*%*%*%*%*%*%*%*%*%*%*%*%*%*%*%*%*
%*%*%*%*%*%*%*%*%*%*%*%*%*%*%*%*%*%*%*%*%*%*%*%*
\section{Summary}\label{sec:Sum}
To summarize, we developed a general and simple formalism to consider the feasibility of contact (degeneracy of energy bands), which can deal with the accidental as well as essential degeneracies. 
The cardinal standpoint of our framework is the feasibility (generalized von-Neumann-Wigner) theorem, $n_{\rm d}=n_{\rm u}-m^2+1-n_{\rm c}$, which provides the number of constraints on the lattice, $n_{\rm c}$, necessary to have a feasible $n_{\rm d}$-dimensional contact with $m$-fold degeneracy by some tuning of $n_{\rm u}$-unknown parameters in multi-band systems.
It enables us to judge {\it without patiently solving the secular equation} which lattice affords feasible contacts.
Primarily, our framework provides a practical procedure to pick up only the degenerate solutions of the secular equation, i.e., only $\bm k$-points with contact, selectively. 
This procedure plays an essential role in the design of Dirac systems, because the Dirac points often appear as an accidental degeneracy at unknown general $\bm k$-points. 
\par
In simpler terms, the usual Dirac points at general $\bm k$-point in two-dimension correspond to the case with $n_{\rm d}=0$ (point contact), $m=2$ (two-fold degeneracy), and $n_u=2$ ($k_1$ and $k_2$), in which case, a single constraint, $n_{\rm c}=1$, is required to fulfill the feasibility condition. 
Such case can be practically explored in many actual crystals. In fact, besides a well known $\alpha$-ET$_2$I$_3$, another 2D organic crystal, (DIEDO)$_2X$ ($X$=Cl,Br), is found to have Dirac points whose space group is P$\bar{1}$, i.e., only inversion symmetry present  (for details of band structure, see Fig.~14 in Ref.~\onlinecite{imakubo06}). 
Even in the spin-dependent cases, the feasible contacts can be found in the same manner in the 2D systems, if some extra constraints besides the space-time inversion symmetry is present; we showed as an application to the LaOFeAs systems that the reflection symmetry works as such extra constraint, and allows for the emergence of Dirac points in the spin-density-wave state. 
\par
It is noteworthy that in three-dimension, the feasibility condition for the point contact at general $\bm k$-point is fulfilled in the absence of constraints: Eq.~(\ref{eq:n_d}) holds for $(n_{\rm d},n_{\rm u},m,n_{\rm c})=(0,3,2,0)$. 
Therefore, the 2D lattice is favorable for the usual crystals which often have at least one symmetry such as inversion. 
\par
\acknowledgments
This work is supported by Grant-in-Aid for Scientific Research (No. 21740231, 20104010, 21110522, 19740218, 22014014) from the Ministry of Education, Science, Sports and Culture of Japan. 
\par
%*%*%*%*%*%*%*%*%*%*%*%*%*%*%*%*%*%*%*%*%*%*%*%*
\appendix
\section{Derivation of Eq.~(\ref{eq:eqf1})}\label{sec:app1}
Expanding $E_0(\bm k,\epsilon)-\epsilon_0$ and $\bm R(\bm k,\epsilon)$ within the linear order of $\delta\bm k=\bm k-\bm k_0$ and $\delta\epsilon=\epsilon-\epsilon_0$, we obtain the self-consistent equation for $\delta\epsilon_\pm=\epsilon_\pm(\bm k)-\epsilon_0$ as
\begin{align}
\delta\epsilon_\pm=&\left(\delta\bm k\cdot\nabla_{\bm k}E_0+\delta\epsilon_\pm\nabla_\epsilon E_0\right)\notag\\
&\ \ \ \pm\left|\delta k_1\nabla_{k_1}\bm R+\delta k_2\nabla_{k_2}\bm R+\delta \epsilon_\pm\nabla_{\epsilon}\bm R\right|,
\label{eq:self_xi}
\end{align}
where all derivatives should be evaluated at $\bm k=\bm k_0$ and $\epsilon=\epsilon_0$.
On the other hand, the self-consistent equation for $\delta \xi_j\equiv \xi_j(\bm k)-\epsilon_0$ reads
\begin{equation}
\delta \xi_j=\delta\bm k\cdot\nabla_{\bm k}E_0+\delta\xi_j\nabla_\epsilon E_0,
\end{equation}
which gives
\begin{equation}
\delta\xi_j=\delta\bm k\cdot\nabla_{\bm k}\xi_j=(1-B)^{-1}\delta\bm k\cdot\nabla_{\bm k}E_0
\label{eq:delta_xi}
\end{equation}
with $B=\nabla_\epsilon E_0$.
The linearization of the relation, $\bm R_j(\bm k)=\bm R(\bm k,\xi_j(\bm k))$, also gives
\begin{equation}
\bm X_j\delta k_1+\bm Y_j\delta k_2=\delta k_1\nabla_{k_1}\bm R+\delta k_2\nabla_{k_2}\bm R+\delta\xi_j\nabla_\epsilon\bm R,
\label{eq:Rj_and_R}
\end{equation}
where $\bm X_j=\nabla_{k_1}\bm R_j$ and $\bm Y_j=\nabla_{k_2}\bm R_j$.
Thus, the self-consistent equation (\ref{eq:self_xi}) is rewritten as
\begin{equation}
(1-B)\Delta_\pm=\pm\left|\bm X_j\delta k_1+\bm Y_j\delta k_2+\bm C\Delta_\pm\right|,
\end{equation}
with $\Delta_\pm\equiv\delta\epsilon_\pm-\delta\xi_j=\epsilon_\pm(\bm k)-\xi_j(\bm k)$ and $\bm C=\nabla_\epsilon\bm R$.
Its solution reads
\begin{equation}
\Delta_\pm=D^{-1}\left(\bm R_j\cdot\bm C\pm\sqrt{(\bm R_j\cdot \bm C)^2+D\bm R_j^2}\right),
\label{eq:Delta}
\end{equation}
where $\bm R_j(\bm k)$ is expanded as $\bm R_j=\bm X_j\delta k_1+\bm Y_j\delta k_2$, and $D=(1-B)^2-\bm C^2$ is introduced.
Then, we finally obtain Eq.~(\ref{eq:eqf1}), using $\epsilon_\pm(\bm k)=\epsilon_0+\delta\xi_j+\Delta_\pm$ and $\delta\xi_j=\delta\bm k\cdot\nabla_{\bm k}\xi_j$.
It is noteworthy that we do not need explicit functional form of $\xi_j(\bm k)$ to evaluate $\bm X_j$, $\bm Y_j$, and $\nabla_{\bm k}\xi_j$. 
Actually, they are given as Eq.~(\ref{eq:derivs}), owing to Eq.~(\ref{eq:delta_xi}) and (\ref{eq:Rj_and_R}).
\par
Now, let us introduce the $2\times 2$ matrix,
\begin{align}
\hat M
&=(1-B)\hat I-\bm C\cdot\hat{\bm\sigma}\notag\\
&=\hat I-\nabla_\epsilon\hat H^{\rm (eff)}\notag\\
&=\hat I+\hat H_{\rm AB}(\epsilon_0-\hat H_{\rm BB})^{-2}\hat H_{\rm BA}.
\end{align}
Its eigenvalues are not less than one, since it satisfy
\begin{equation}
\langle\phi|\hat M|\phi\rangle=\bigl|\!\bigl||\phi\rangle\bigr|\!\bigr|^2+\bigl|\!\bigl|(\epsilon_0-\hat H_{\rm BB})^{-1}\hat H_{\rm BA}|\phi\rangle\bigr|\!\bigr|^2\ge 1,
\end{equation}
for any normalized $|\phi\rangle\in S_{\rm A}$.
Thus, we obtain
\begin{equation}
D=(1-B)^2-\bm C^2=\text{det}\,\hat M\ge 1,
\end{equation}
and see that the value inside square root never become negative in Eq.~(\ref{eq:Delta}).
Also, the factor $(1-B)^{-1}$ never diverges in Eq.~(\ref{eq:derivs}), due to the inequality, $(1-B)^2\ge 1$.
\par
\vspace{4mm}
%*%*%*%*%*%*%*%*%*%*%*%*%*%*%*%*%*%*%*%*%*%*%*%*
\section{Spin-dependent cases at $\bm k=\bm G/2$}\label{sec:app2}
In Ref.~\onlinecite{murakami07}, the spin-dependent cases are considered in detail at the special $\bm k$-points, $\bm k=\bm G/2$, where we can expect larger number of constraints than at the general $\bm k$-points.
In fact, $\hat H^{\rm (eff)}(\bm k,\epsilon)$ is invariant under the time-reversal and inversion separately there, which leads to
\begin{equation}
\left\{
\begin{array}{l}
\hat h^{\rm (eff)}_{\uparrow\uparrow}(\bm G/2,\epsilon)=\hat h^{{\rm (eff)}*}_{\downarrow\downarrow}(\bm G/2,\epsilon)\\
\hat h^{\rm (eff)}_{\uparrow\downarrow}(\bm G/2,\epsilon)=-\hat h^{{\rm (eff)}T}_{\uparrow\downarrow}(\bm G/2,\epsilon),
\end{array}
\right.
\label{eq:T-rev-spin}
\end{equation}
and
\begin{equation}
\hat h^{\rm (eff)}_{\sigma\sigma'}(\bm G/2,\epsilon)=\hat U\hat h^{\rm (eff)}_{\sigma\sigma'}(\bm G/2,\epsilon)\hat U^{-1},\label{eq:inv-spin}
\end{equation}
respectively.
In the case of  $\bm\omega=\bm 0$ (i.e., $\hat U=e^{i\phi}\hat I$), the number of constraints is unchanged.
Otherwise, it is increased up to fourteen, since we obtain $\bm s^{(1)}\cdot\bm R=\bm s^{(2)}\cdot\bm R=0$ and $Z_1=Z_2=0$ from Eqs.~(\ref{eq:T-rev-spin}) and (\ref{eq:inv-spin}), where $\bm s^{(1)}$ and $\bm s^{(2)}$ are the linearly independent vectors perpendicular to $\bm\omega$.
In the latter case, a single extra constraint is necessary to find a feasible point contact on these special $\bm k$-points ($n_{\rm u}=0$), because Eq.~(\ref{eq:n_d}) gives $n_{\rm c}=m^2-1-n_{\rm u}=15$.
However, such a feasible point contact is out of our main interest, because it is no longer accidental but essential. 
\par
It should be noted that our interest differs from that of Ref.~\onlinecite{murakami07}.
Actually, the point contacts studied in Ref.~\onlinecite{murakami07} is {\it unfeasible} in our context.
It needs the fine-tuning of a single lattice parameter for its realization, or equivalently, an extra constraint for its feasibility. 
In terms of our formalism, they consider the point contact under the condition, $n_{\rm u}-m^2+1+n_{\rm c}=-1$.
\par
\section{Kramers degeneracy}\label{sec:app3}
In this appendix, we mention how the Kramers degeneracy is derived from the space-time inversion symmetry using our formalism given in \S\ref{sec:GvNW-theorem}.
Focusing on a single atomic orbitals, we obtain $2\times 2$ effective Hamiltonian matrix, $\hat H^{\rm (eff)}$, with the same form as Eq.~(\ref{eq:heff-spin}), whereas $\hat h^{\rm (eff)}_{\sigma\sigma}$ are no longer matrices but scalars.
We can adopt the consideration for the site-centered inversion, and obtain Eq.~(\ref{eq:site-centered-spin}), which results in
\begin{equation}
\hat h^{\rm (eff)}_{\uparrow\uparrow}=\hat h^{\rm (eff)}_{\downarrow\downarrow},\ \ \hat h^{\rm (eff)}_{\uparrow\downarrow}=-\hat h^{\rm (eff)}_{\uparrow\downarrow}=0,
\end{equation}
because $\hat h^{\rm (eff)}_{\sigma\sigma'}$ are scalars.
Thus, the two-fold essential degeneracy takes place at every $\bm k$-point, since the effective Hamiltonian always becomes a $2\times 2$ scalar matrix, $\hat H^{\rm (eff)}=\hat h^{\rm (eff)}_{\uparrow\uparrow}\hat I$.
\par
%*%*%*%*%*%*%*%*%*%*%*%*%*%*%*%*%*%*%*%*%*%*%*%*
\end{document}